# Evolution of public funding for collaborative health research towards higher-level patient-oriented research

## *A comparison of the European Union Framework Programmes to the program funding by the United States National Institutes of Health*

David Fajardo-Ortiz. Department of Cardiovascular Sciences, KU Leuven, Leuven, 3000, Belgium. Faculty of Medicine, National Autonomous University of Mexico (UNAM), Ciudad Universitaria 04510, Mexico. ORCID: 0000-0002-7657-7514

Bart Thijs. Department of Management, Strategy and Innovation, Faculty of Economics and Business, KU Leuven, Leuven 3000, Belgium. ORCID: 0000-0003-0446-8332

Wolfgang Glänzel. Department of Management, Strategy and Innovation, Faculty of Economics and Business, KU Leuven, Leuven 3000, Belgium. . ORCID: 0000-0001-7529-5198

Karin R. Sipido. Department of Cardiovascular Sciences, KU Leuven, Leuven 3000, Belgium. ORCID: 0000-0001-9609-5507

Corresponding Author: Karin Sipido

Karin.Sipido@kuleuven.be

Division of Experimental Cardiology

Department of Cardiovascular Sciences

KU Leuven

Campus Gasthuisberg

Herestraat 49 – bus 911

B-3000 Leuven

Belgium

phone +32-16-330815

## Abstract

Public research funding agencies increasingly seek to steer health research toward higher levels of translation and societal relevance. Yet it remains unclear to what extent such policy shifts are effectively implemented and reflected in funded projects and scientific outputs. This study examines evolution and changes in the orientation of health research portfolios since 2008 within European funding (Framework Programmes FP7 and Horizon 2020 funding for collaborative health research, FP-HR, and ERC Life Sciences grants), in comparison to NIH funding for collaborative research (P01, U01, and UM1). Using large-scale text analysis and supervised classification, we analyze both project descriptions and the associated scientific publications.

At the project level, the EU FP-HR show pronounced shifts toward population-level, diagnostic, and health systems–oriented research, whereas investigator-driven ERC life sciences, NIH P01 and U01, display greater stability with a predominance of basic biomedical research. Publication-level analyses reveal more moderate changes, with basic biomedical research remaining a central component including in EU FP-HR, indicating partial translation of funding priorities into outputs.

By jointly analyzing projects and publications, this study identifies and distinguishes between changes in funder expectations and realized research trajectories, highlighting how strategic funding shapes research portfolios within enduring epistemic and institutional constraints.

**Keywords:** Public funding of science; Funding priorities; FP7; Horizon 2020; Horizon Europe; Research Project with Complex Structure Cooperative Agreement

# 1 Introduction

Expectations of societal benefit is the rationale for public investment in research. The impact of science and research at public institutions on technological development during World War II, inspired and motivated the launch of the National Science Foundation, NSF, in the United States, US, one of the oldest governmental funding agencies (Bush, 1945, 1947). Stimulating frontier science, driven by scientific curiosity and opportunity, was the aim and excellence of the proposed research the lead awarding criteria (Shaw, 2022). To this day, the model of science-led priorities is still predominant in most governmental funding programs of academic research. However, under pressure for enhanced and faster return for economic growth, new models have emerged, such as bringing the private sector together with academia in public-private partnerships (Arnold & Barker, 2022). Other funding mechanisms define strategic priorities, with more explicit endpoints and deliverables, as set out in the concept of missions (Mazzucato, 2018). Societal impact, contributing to the sustainable development goals, has become a major driver for public investment in research (Kastrinos & Weber, 2020). Unlike research funding guided by the interest of researchers and excellence of their proposals, this strategic funding seeks to provide the necessary knowledge to meet society's needs for scientific knowledge in the face of the challenges it faces.

Health and biomedical research in particular carry the expectation of a healthier society as return on investment, with economic perspectives for a healthy workforce and a thriving healthcare industry (Caulfield, 2010; Nathan et al., 2001). In the early 2000’s, a noted lack of progress from fundamental biomedical research and discoveries to new treatment, the so-called translational gap (Aarden et al., 2021), has stimulated strategic funding programs in health research (Llewellyn et al., 2020; Reed et al., 2012). The recent Cancer Moonshot (Singer, 2022) in the US and the Mission: Cancer in the European Union (Lawler et al., 2021), are paradigmatic examples of deliverable-oriented funding, and typically engage actors from the so-called quadruple helix, i.e. academia, the private sector or industry, government and society, in this case patients (Miller et al., 2018). More recently, the Covid-19 pandemic and the threat of emerging infectious diseases has further stimulated targeted investment (OECD, 2021). The strategy of priority funding to areas of need, is complemented by two other strategies to stimulate translation towards better patient care. The first is stimulation of collaboration in multi- and cross-disciplinary research and seeking critical mass, in particular for clinical research (Al et al., 2023; Demotes-Mainard & Ohmann, 2005). The second is incentivizing research aiming for implementation of knowledge, such as clinical and diagnostics research and research toward health policies and management (Neta et al., 2015; Roberts et al., 2019).

At EU level, investment in R&D is closely linked to the overall EU political priorities. Amendments to the EU Treaty, starting in 1986, progressively consolidated the investment in research and innovation by the EU and set out the expectation for economic return. In a 7-year cycle, the EU Regulations on the Framework Programmes, FPs, funding research and innovation, have since laid down defined strategies (Kim & Yoo, 2019; Rodríguez et al., 2013). The content of the FPs is an example of the complexity of the political process of setting priorities (Ludwig et al., 2022).

A recent study commissioned by the Foresight Unit STOA of the European Parliament (Sipido et al., 2022) highlighted the importance of funding for health research through the strategic targeted programs within the FPs. Although in terms of budget, the EU FPs contribute but a fraction of the total health research investment in Europe, the FP funding is unique in its policy-driven strategy and support for international collaborative cross-border research. Also, the EU is the second largest funder of health research in the world, after the United States National Institutes of Health (NIH) (Sipido et al., 2022; Viergever & Hendriks, 2016). Of note, the FPs also fund frontier science under a researcher-driven model through the European Research Council (ERC) (König, 2017; Luukkonen, 2014). Within the NIH on the other hand, the vast majority of health research funding is under an investigator-initiated competitions model though the NIH reserve some resources to fund top-down science (Myers, 2020). Supplement 1 further details funding policies and strategic planning of the EU FPs and the NIH.

A common driver for EU and NIH has been the ambition for more translation and implementation of findings for better health (Aarden et al., 2021), but whether there is convergence in the evolution of the type of health research funded by these lead agencies with their different policies is unknown.

A substantial body of research has examined how public funding for health research is distributed across disease areas and how these allocations relate to indicators such as disease burden, health outcomes, or returns on investment (Ballreich et al., 2021; Bowen & Casadevall, 2015; Gillum et al., 2011; Gross Cary et al., 1999; Madsen & Andersen, 2024). Related work has also explored how targeted funding shapes the direction of scientific activity within specific disease domains and the extent to which spillovers or unexpected research trajectories emerge (Coburn et al., 2024; Yaqub et al., 2024). While this literature has provided important insights into funding priorities and their consequences, it has focused primarily on thematic or disease-level allocations and, in many cases, on single funding systems or static time periods. By contrast, the present study adopts a comparative, portfolio-level perspective across multiple funding mechanisms and programme periods, explicitly distinguishing between the orientation of funded projects and the content of resulting scientific publications. This approach allows us to examine not only how funding priorities change, but also how such changes are partially realized in research outputs.

Therefore, the aim of our study is to assess whether, and to what extent, stated policy ambitions for greater translation, implementation, and patient-oriented research are reflected in the actual content of publicly funded collaborative health research projects and their output. We address this aim through a comparative analysis of funding portfolios from the European Union Framework Programmes (FPs) and selected funding mechanisms of the United States National Institutes of Health (NIH).

The study is guided by three interrelated research questions. First, we ask whether successive EU Framework Programmes have effectively translated evolving policy priorities into a reorientation of funded health research projects towards higher levels of research, closer to clinical application, population health, and health systems implementation. Second, we examine whether a comparable shift can be observed within NIH collaborative funding mechanisms, despite their predominantly investigator-driven and bottom-up nature. Third, we assess whether differences in funding models—periodically redesigned, call-based programmes in the EU versus more stable, investigator-initiated mechanisms at the NIH—are associated with different capacities to adjust research portfolios over time in response to policy objectives.

In this perspective, the comparison between EU and NIH funding is not intended as a direct evaluation of national performance, but rather as an analytical contrast between two structurally distinct models of public research funding. This contrast allows us to examine how institutional design and governance of funding instruments shape the translation of policy goals into funded research content.

In this article, the distinction between "bottom-up" and "top-down" funding mechanisms is used as an analytical simplification to capture differences in portfolio-level governance and responsiveness to policy priorities of the programmes selected for the study. The terms do not imply a strict dichotomy at the level of individual projects, nor do they suggest incompatibility between strategic orientation and curiosity-driven research in the broader funding landscape. Rather, they serve as heuristic categories to facilitate comparison of funding logics across the programmes and time periods under study.

For our study, we selected two periods based on the life cycle of the EU FPs, i.e. 2008-2014 (FP7) and 2015-2021 (Horizon 2020, H2020), complemented with the most recent data up to 2023 funded under the present FP Horizon Europe. Using text-based analysis of the project abstracts, we asked whether health research projects were primarily in the area of basic biomedical research or beyond, seeking translation and implementation. We identified the clinical-therapeutic level, as well as the levels of diagnostics, prognosis and screening, population and risk factor research, and health management and policies. We also gathered evidence on project funding for some major challenging health problems, to inform on prioritization.

By combining large-scale text-based classification of funded projects with an analysis of their associated research outputs, the study moves beyond descriptive trend analysis and provides empirical evidence on the alignment between policy intent, funding instruments, and knowledge production in health research.

# 2 Methods

## 2.1 Selection criteria of EU and NIH funding programs for analysis

The funding program under FP7's Pillar Health (further referred to as FP7-HR), under Horizon 2020 Societal Challenge Health (further referred to as H2020-HR), and under Horizon Europe Cluster Health, finance collaborative projects that respond to strategically defined calls. We identified three funding programs of NIH that share characteristics with these EU programs. The U01 projects, which are defined as "Research Projects-Cooperation Agreements," adhere to the bottom-up scientist-driven funding model, but with a dimension of purpose and goal, and explicitly expected cooperation: '.. *support a discrete, specified, circumscribed project to be performed by the named investigator(s) in an area representing his or her specific interest and competencies*.' For our analysis, we selected only those U01 projects with two or more principal investigators given our goal of comparing FPs to NIH funding of collaborative projects. P01, which are defined as "Research Program Projects," also follow the researcher-driven funding model and include various research teams led by independent researchers working on interdependent projects, but do not stipulate the cooperation that is expected in the U01. UM1, which are defined as "Research Project with a Complex Structure Cooperation Agreements," are in an intermediate position between the strategic financing model and bottom-up financing for research, since depending on the center within the NIH, these projects can originate as response to specific calls or proposals from researchers. The main characteristic is that they are large-scale projects with highly complex structures such as clinical networks or consortiums. It is important to mention that these three NIH funding streams (U01, P01 and UM1) are for extramural research projects, and the distribution of these funds is under the administration of the 27 institutes and centers of NIH (Supplementary information 1). For the EU funding through the FPs, we included in the analysis funding through the European Research Council, ERC, a funding program without topic- defined calls. ERC is a paradigmatic example of the researcher-driven funding model, in which a principal investigator proposes a research project to compete for funding with academic excellence being the only criteria. We incorporate these projects into the analysis because they are the result of a policy decision for investment in excellence-driven research. Thereby they serve as a reference data set without strategic influence on topic selection, only driven by scientific excellence and exemplary for fundamental frontier research.

In total, 26,510 projects were analysed, of which 3,704 ERC-LS projects (1,926 and 2,358 in the first and second period respectively), 1,008 under FP7-HR, 1,216 under H2020-HR, 9,328 NIH-U01 projects with two or more principal investigators (2,040 and 7,284 for each period respectively), 8,891 P01 projects (5,405 plus 3,280 for both periods), and 2,363 UM1 projects. Figure 1 illustrates the associated funding amounts for the two periods under study.

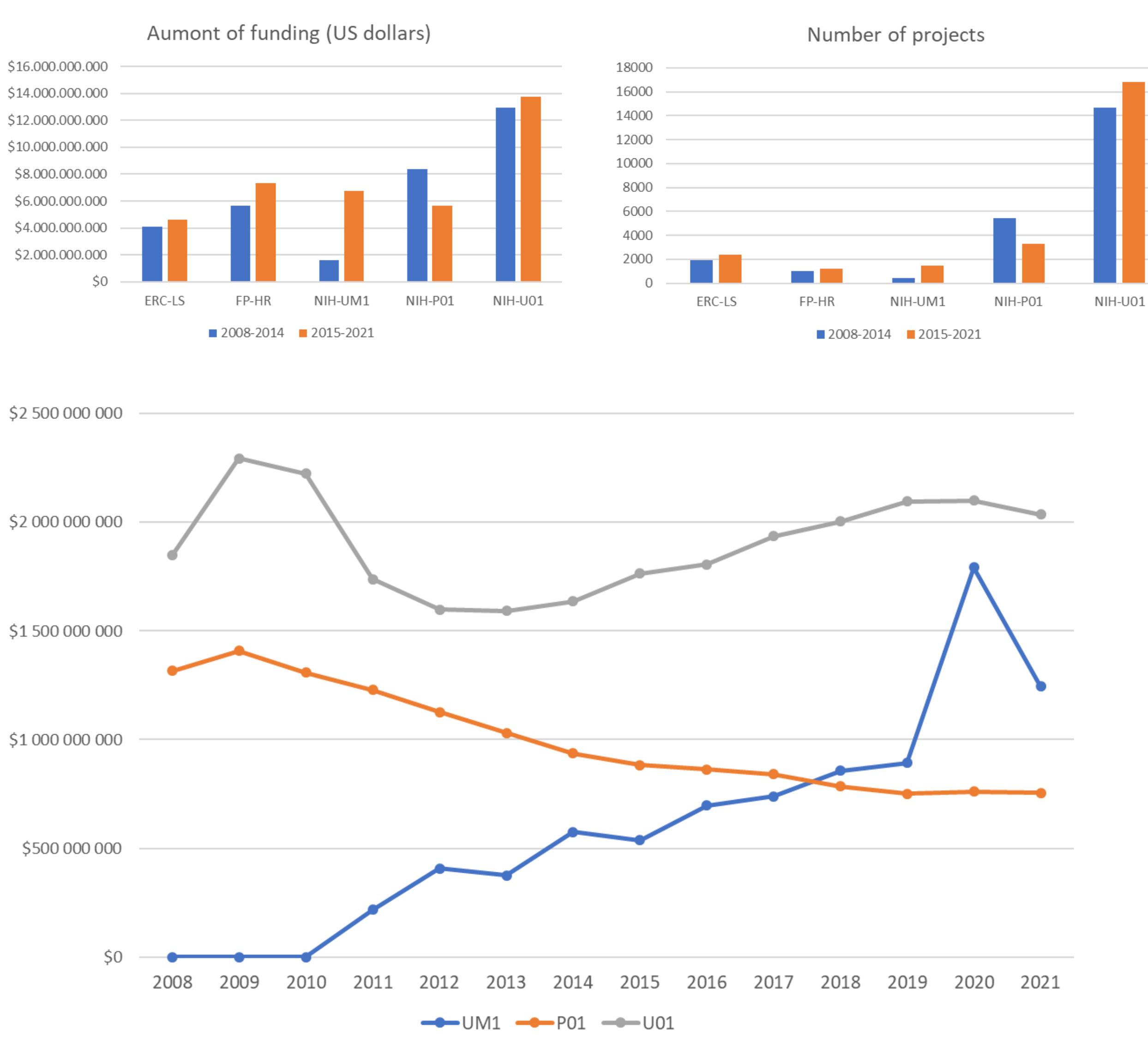

**Figure 1.** *Level of funding and number of projects funded under the programs included in the study.*

In addition, 344 Horizon Europe Cluster Health projects were analyzed as well as 458 ERC-LS projects, to extend insight in the evolution of EU funding over 3 generations (FP7, H2020 and Horizon Europe up to 2023).

## 2.2 Data sources for project funding and content

As sources of information for EU funding, we used the CORDIS database (https://cordis.europa.eu/ ) to extract information on health research projects financed by FP7, Horizon 2020 and Horizon Europe, and the datahub of projects funded by the ERC (https://erc.easme-web.eu/). We used

RePORTER (https://reporter.nih.gov/ ) to extract information on health research projects funded by the NIH via U01 (Research Project-Cooperative Agreements), P01 (Research Program Projects) and UM1 (Research Project with Complex Structure Cooperative Agreement).

Except for the projects financed by the NIH via UM1, which is a relatively recent program (the first projects financed appear in fiscal year 2011), projects were selected for the periods 2008–2014 and 2015–2021. For the European Union, these periods correspond to successive Framework Programme (FP) generations, which provide a natural temporal structure for analysing shifts in funding orientation. Indeed, Framework Programmes are designed as multi-year strategic cycles reflecting changing views on and expectations for investment, justifying the choice for the programme period as the primary temporal unit of analysis. Although adjustments to funding priorities may occur within programme periods, these changes typically concern specific subtopics or instruments rather than the overarching thematic orientation and research-level configuration that are the focus of this study. Using programme generations as analytical units therefore enables systematic longitudinal comparison while acknowledging that finer-grained within-programme dynamics may not be fully captured.

The three databases used provide a summary description (title, abstract) of the financed projects as well as metadata on the identity of the principal investigator, the beneficiary institutions and the amount of support. For the content analysis of the research projects, we used titles and abstracts' text.

## 2.3 Definition of levels in health research.

In the present study, we use five levels of health research, which are previously well defined in the scientific literature (Fajardo & Castano, 2016; Hoffman et al., 2012; Nederbragt, 2000; Pratt et al., 2020): **Basic biomedical research** consists of the study of the pathophysiological mechanisms that explain the disease at the sub individual level. This level includes the formulation of potential pharmaceutical or biotechnological therapies that are tested in vitro and animal models. **Clinical-therapeutic research** is related to the development and evaluation of new treatments against diseases, that is, the development of disease interventions at the individual level. This level includes clinical trials phases 1 to 4. **Research on diagnosis and screening** is also clinical research but aimed to the identification of disease in individuals to guide treatment and patient management**. Research on risk factors in the population**, as well as basic biomedical research, seeks to explain the disease, but at a collective level. This research includes the search for genetic, environmental and social determinants or factors of the disease. Finally**, research on health management and policies**, in a similar way to clinical research, seeks to intervene in health problems but through the design, implementation and evaluation of organizational efforts.

The classification scheme was based on an explicit, rule-based conceptual framework grounded in established distinctions in health research. Specifically, it was structured along two analytical dimensions. The first dimension captures the unit of analysis of research activity, ranging from sub-individual (e.g., molecular or cellular mechanisms), to individual (clinical research), to population-level inquiry (epidemiology and public health), and to systems-level or governance-oriented research (health policy and management). The second dimension reflects the orientation of research from basic to applied. These dimensions provided a consistent conceptual grid to guide expert annotation of the training corpus, ensuring that classification decisions reflected underlying research orientation rather than isolated keywords. The same framework subsequently informed the supervised machine-learning model, aligning automated classification with the conceptual structure used in manual coding.

## 2.4 Text mining tools and algorithms

The content of the project descriptions was analyzed in three complementary workflows as illustrated in Figure 2.

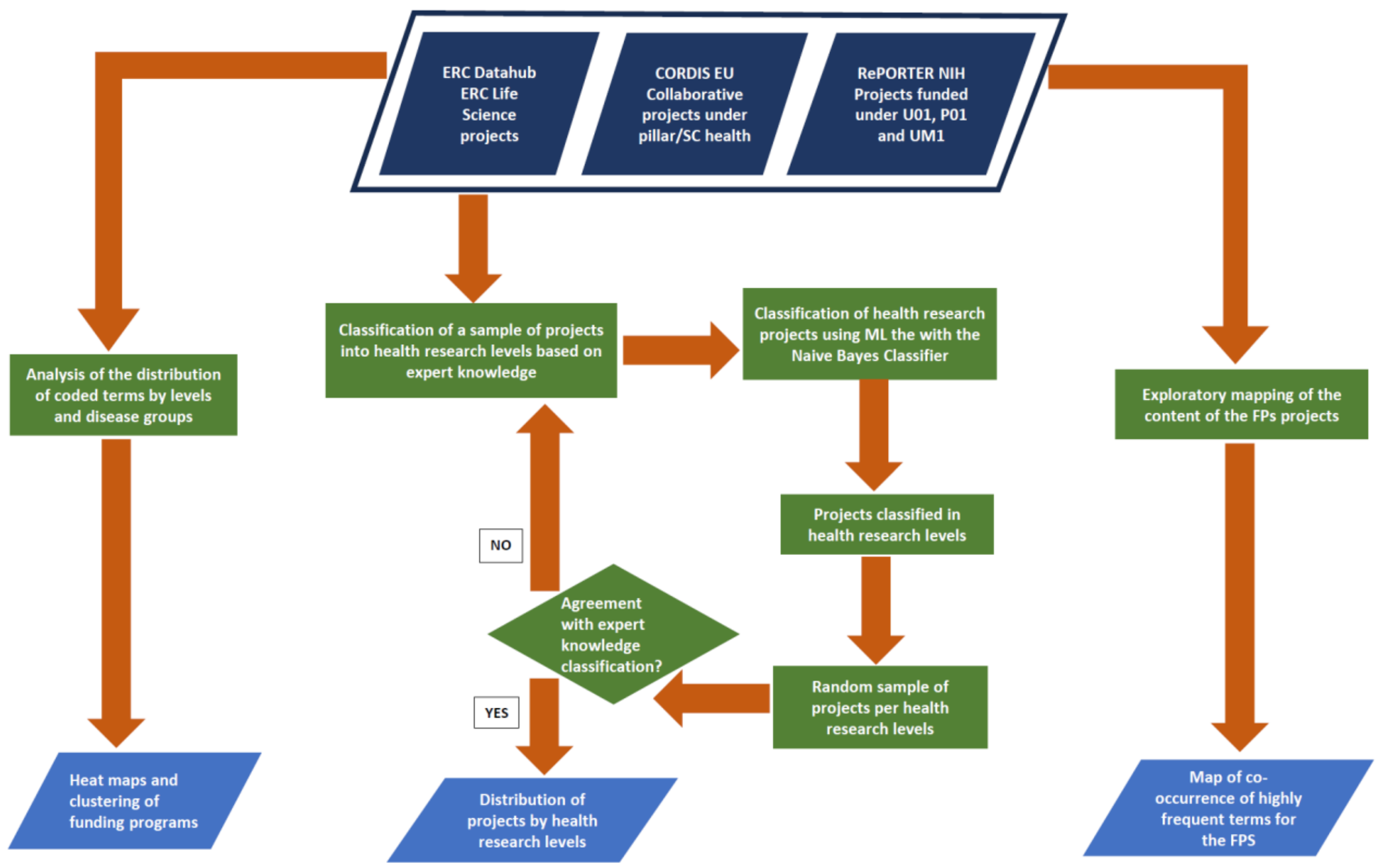

**Figure 2.** *Diagram of the workflow of the analysis of information on projects funded by NIH and FP HR. At the top are source data, green indicates steps of analysis and blue resulting data. * Levels of health research: basic biomedical, clinical therapeutic, diagnostics-prognosis-screening, population and risk factors, and management and health policies.*

### 2.4.1 Distribution analysis of coded terms by research levels and disease groups

The first workflow consists of an analysis of the distribution of highly frequent terms coded by research levels resulting in a heatmap and clustering of funding sources according to research levels. To this end, the content (title and abstract) of the description of the funded projects was processed and analyzed using KH Coder, a powerful tool for statistical content analysis based on R, to identify the main terms associated with each program (Higuchi, 2016). The results of the exploratory content analysis identified a set of highly frequent key terms that were codified into five predefined levels of research: basic biomedical, clinical therapeutic, diagnostics-prognosis-screening, population and risk factor research, and health policies and management. We then performed a hierarchical cluster analysis of the funding mechanisms based on the distribution of coded terms across five levels of research, and present results in heatmap.

Using the same tools, we generated a heatmap and clustering according to disease groups: cancer, infectious diseases and cardiometabolic diseases. Due to the diversity and dispersion of terms associated with disease groups, instead of searching among highly frequent terms, we took advantage of the extensive specialized vocabulary contained in the Medical Subject Headings (MeSH) categories and their input terms, discarding general terms such as cell, tissue or disease and avoiding the repetition of words in the coding. In the case of neoplasms, we use all the entry terms of the "neoplasm" category and all the subcategories and entry terms of "neoplasms by histologic type" and "neoplastic processes." For cardiometabolic diseases we used the MeSH categories "glucose metabolism disorders, "lipid metabolism disorders," "metabolic syndrome," "overnutrition," "heart diseases," and "vascular diseases." Finally, for infectious diseases we used the MeSH categories "bacterial infections and mycoses," "parasitic diseases," and "virus diseases".

### 2.4.2 Classification and distribution of project funding according to levels of research

The second workflow consists of iterative classification of projects that uses machine learning tools, which were trained based on expert knowledge, to obtain a distribution of projects according to the different research levels within each funding program. In a first round, a sample of 10 representative projects from each research level (basic biomedical, clinical therapeutic, diagnostics-prognosis-screening, population and risk factor research, and health policies and management), 50 in all, was classified based on the expert knowledge of the authors. This sample was then used as starting point to build a model for project classification using KH Coder's Naive Bayes Classifier, a machine learning tool (Iwata et al., 2021). A model, in plain terms, is a set of terms, each with a score related to the probability of being associated with one category or another, built from the examples manually classified based on the expert knowledge of the authors. As the number of examples of manually classified projects increases, a refinement of the capacity of the Bayes classifier model to discern correctly to which level of research a project would belong is expected, until saturation is reached. Thus, when in the final round the number of projects classified manually for the construction of the model was increased to 100, the distribution of projects classified by research levels among the different funding channels no longer changed significantly.

KH Coder automatically reclassified the 100 projects used as input for building the model to perform cross-validation; 80 of the projects used as input for the model were correctly classified. At the end of the final round of classification, we extracted a random sample out of the 26,510 projects analyzed to evaluate the classification performed, with 82% agreement between the experts and KH Coder. Once having a satisfactory classification, an analysis of the distribution of the projects classified in the five levels of health research was carried out, among the different financing programs for the two periods considered (2008-2014 and 2015-2021).

Alternative classification approaches exist, including the Medical Text Indexer (MTI) and more recent large language model–based methods. MTI, for example, has been shown to achieve substantial recall across document types such as grants and patents, although precision varies depending on document characteristics and aggregation level (Moore et al., 2024). Given the scale of our corpus and the need for interpretability and reproducibility in a comparative, longitudinal setting, we opted for a Naive Bayes classifier as a well-established baseline method that allows transparent inspection of term–category associations. The study does not claim methodological optimality; rather, it prioritizes consistency and conceptual alignment across datasets and programme periods.

Health research projects frequently encompass elements spanning multiple research levels, reflecting the intrinsically cross-level nature of biomedical and public health inquiry. Accordingly, classification was based on the dominant research orientation identified in the text rather than on exclusivity of level-specific terminology. In cases where multiple levels were represented, the assignment reflected the most salient and substantively emphasized orientation. Highly ambiguous cases were rare, and sensitivity checks using alternative threshold criteria did not materially alter aggregate programme-level distributions, supporting the robustness of this approach.

### 2.4.3 *Exploratory mapping of the content of the FPs projects*

A third analysis was added in order to provide greater context to the evolution of the content of the health cluster of the framework programs, FP7, H2020 and Horizon Europe. Here, we performed an exploratory content analysis to determine which high frequent terms are most associated with one of the three different generations of the framework program.

For text processing using KH Coder we selected the lemmatization option, which ensures that the output word is an existing normalized form of the word that can be found in the dictionary, that is, the "lemma." The lemmatization performed by KH Coder uses the “Stanford POS tagger” that allows distinguishing the parts of speech (nouns, pronouns, proper names, adjectives, verbs, etc.). The co-occurrence networks were elaborated using only nouns, which facilitated the interpretation and avoided the duplicate appearance of terms that are spelled the same but are actually different parts of speech (such as the verb "challenge" and the noun "challenge").

The results were presented through a bimodal network of co-occurrence of terms and generations of the FPs. For better visualization and interpretation of the results, the edges of the network were filtered at 60 with the strongest Jaccard association coefficient.

Additional analysis of co-occurrence maps within the NIH programs and underlying methods are presented in the Supplementary Information. For this additional analysis, the co-occurrence network clustering using KH Coder was compared with four different alternative algorithms: 1) Leiden, 2) Louvain, 3) principal eigenvector, and 4) edge-betweenness communities (Supplementary Information).

These analyses are complemented by a parallel examination of scientific publications resulting from funded projects, enabling a comparison between project orientations and research outputs.

## 2.5 Publication data and analytical strategy

To assess whether changes in funding priorities observed at the project level are reflected in research outputs, we complemented the project-level analysis with a systematic analysis of scientific publications resulting from funded projects. Figure 3 summarizes the methodological workflow applied to publication data across funding mechanisms and programme periods. In line with the overall design of the study, the unit of analysis is not individual publications or projects, but funding mechanisms, allowing for a comparative assessment of how different funding models translate policy priorities into research outputs.

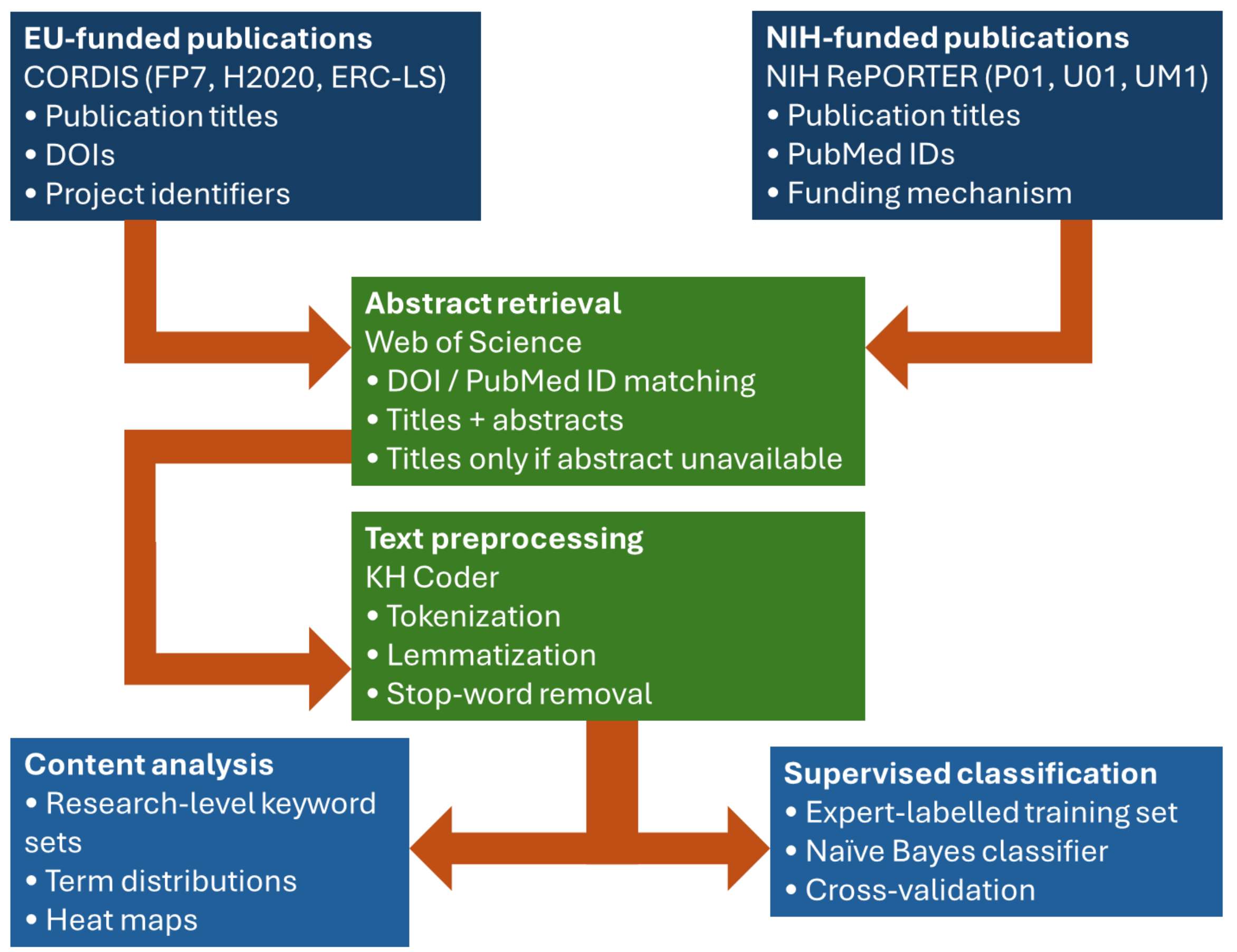

**Figure 3. Methodological workflow for the analysis of publications resulting from funded projects.** *The figure summarizes the analytical steps applied to scientific publications associated with EU Framework Programmes, ERC Life Sciences grants, and NIH funding mechanisms. Publication metadata were retrieved from CORDIS and NIH RePORTER and enriched with abstracts obtained from the Web of Science using DOIs or PubMed identifiers. Texts were processed using the same text-mining procedures as for project descriptions, including lemmatization and keyword-based content analysis. Publications were subsequently classified by research level using a supervised classifier trained and validated on an expert-labelled corpus.*

### *2.5.1 Data sources and linkage to funded projects*

Information on publications associated with EU-funded projects was obtained from the CORDIS database, which provides structured metadata on project outputs, including publication titles and Digital Object Identifiers (DOIs). In the case of the EU Framework Programmes (FP7 and H2020) and ERC Life Sciences (ERC-LS), publications can be directly linked to funded projects through unique project identifiers recorded in the database.

For NIH-funded research, publication data were retrieved from the NIH RePORTER database for projects funded through the P01, U01, and UM1 mechanisms. In contrast to CORDIS, RePORTER does not allow for a direct temporal or component-level linkage between individual publications and specific project years, reflecting the cumulative and non-linear nature of knowledge production in biomedical research. As explicitly stated in the RePORTER documentation, publications are associated with parent projects rather than with specific sub-projects or fiscal years, and publication timelines may extend beyond the formal duration of funding.

To address this limitation, we adopted a funding-mechanism–level approach. For each NIH funding mechanism, we extracted all publications acknowledging support from P01, U01, or UM1 grants and published within defined time windows corresponding to two consecutive programme periods. Specifically, publications published between 2010 and 2016 were associated with projects awarded between 2008 and 2014, while publications published between 2017 and 2023 were associated with projects awarded between 2015 and 2021. This approach allows for a conservative lag between project start and publication while maintaining comparability across funding mechanisms.

#### 2.5.2 Retrieval of publication abstracts

For both CORDIS and RePORTER records, the initial metadata typically include only publication titles. To enable content-based analysis, publication abstracts were retrieved using bibliographic identifiers. For EU-funded publications, DOIs were used to identify records in the Web of Science (WoS). For NIH-funded publications, PubMed Identifiers (PMIDs) provided in RePORTER were used to retrieve corresponding records in WoS. In the small number of cases where abstracts were unavailable, publication titles were used for text analysis.

#### 2.5.3 Text processing and keyword-based content analysis

Textual data (titles and abstracts) were processed using KH Coder following the same preprocessing pipeline applied to project descriptions. This included tokenization, lemmatization, and the removal of stop words. To ensure comparability between project-level and publication-level analyses, we employed the same sets of keywords associated with each research level (basic biomedical research; clinical therapeutic research; diagnostic, prognostic and screening research; population and risk factor research; and health policy and management).

The distribution of key terms across funding mechanisms and programme periods was analyzed and visualized using hierarchically clustered heat maps, allowing for direct comparison between the thematic orientation of funded projects and that of resulting publications.

### *2.5.4 Classification of publications by research level*

In addition to keyword-based analyses, publications were classified into research levels using the same two-dimensional classification framework applied to projects, distinguishing between basic versus applied research and between individual/sub-individual versus population-level orientations.

To construct and validate the classification model, an expert-based training set was developed. A first researcher identified five sets of 50 publications, each clearly corresponding to one research level, based on manual inspection of abstracts retrieved from the Web of Science. A second researcher independently and blindly classified these publications. Disagreements were minimal, with concordance achieved in all but two cases. From this corpus, a balanced training set of 200 publications (40 per research level) was selected to train a Naïve Bayes classifier implemented in KH Coder.

During model construction, cross-validation was performed within KH Coder, yielding an overall classification accuracy of approximately 95%. The trained model was then applied to the full corpus of publications associated with each funding mechanism and programme period.

The robustness of the analytical approach was assessed and rests on two complementary dimensions. First, we combined keyword-based content analysis with supervised classification, allowing comparison between independently derived thematic patterns; the convergence of results across these approaches supports the stability of the observed trends. Second, the supervised classifier was developed through iterative training and cross-validation procedures, and its performance was evaluated against expert-coded data. The agreement between manual expert classification and automated assignment reached 82% for project descriptions and 95% for scientific publications, indicating a high level of consistency between human judgment and machine-based prediction. Together, these elements support confidence in the reliability of portfolio-level distributions.

## 3 Results

### 3.1 Clustering of funding programs in terms of research levels

The first workflow enabled us to relate terms present in the project descriptions to the five levels of research. We analysed the aggregated presence of biomedical, clinical-therapeutic, diagnostics-prognosis-screening, population and risk factors, and health management and policy within each of the funding programs. We then examined the differences and similarities between the different funding programs using an agglomerative cluster analysis based on the Ward method and Euclidean distance. The results are presented as a heat map with dendrogram, shown in Figure 4.

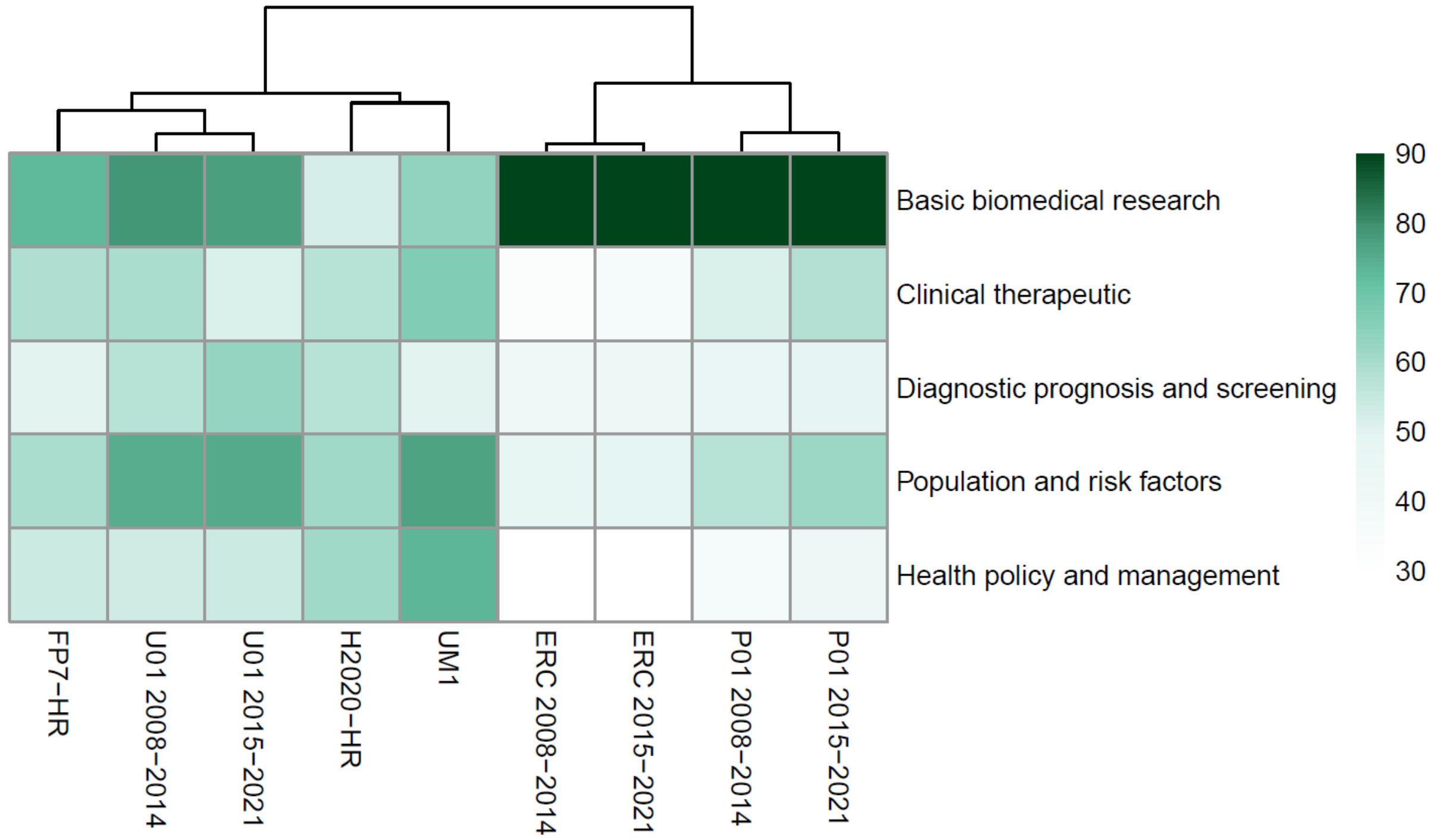

**Figure 4.** *Heat map and clustering of funding programs according to codified domains of health for research projects funded by the NIH and EU FP-HR. The color gradient indicates the percentage of projects that contain in their description terms coded in one of the five research levels.*

The heatmap is organized by the clusters over the different funding channels based on the distribution of the terms coded at the levels of research. At the first, lowest level of the hierarchical clustering of the heat map, three sub-clusters can be identified: NIH-P01, on the right of the map, ERC-LS next to it, and NIH-U01 on the left. Each of these clusters is made up of programs whose research funding profiles are very similar in the two periods analyzed (2008-2014 and 2014-2021), indicating that they changed very little over time. At the next step in the dendrogram, FP7-HR clusters with the NIH-UO1 program, whereas H2020-HR clusters with NIH-UM1. The aggregation of these four programs in two distinct clusters underscores how the FP7 and H2020 health research portfolios are different from each other.

At the third step in the hierarchy, we see how funding mechanisms form two large clusters that differ from each other due to the preponderance of the basic biomedical research level over the others. On the right side of the heatmap there is a cluster formed by projects funded by ERC-LS and NIH-P01, including both periods 2008-2014 and 2015-2021. These four sets of projects are characterized by a high frequency of basic biomedical terms, present in over 90 percent of the projects. Whereas for P01 projects we see some increase in presence of higher research levels (clinical-therapeutic and population) in the second period, this is not the case for ERC. On the left side of the heat map are clustered NIH-U01, NIH-UM1 and the FPs. The projects of these funding channels are characterized by a more equitable distribution of the terms coded at the different levels of research. The two periods analyzed for NIH-U01 are characterized by a high frequency of biomedical and population research terms in over 70% of the projects. FP7-HR also has more than 70% of its projects containing biomedical terms and almost 60% of projects containing terms related to the level of population and risk factors. NIH-UM1 and H2020-HR are characterized by having the lowest percentage of projects with biomedical terms, with a more homogeneous distribution among the different levels of research, consistent with high levels of complexity.

## 3.2 Distribution of projects within funding programs when classified by research levels

Whereas the previous analysis considered the presence of levels of research in the projects measured by terms' frequencies, it was not a hard classification and did not allow analysis of a more quantitative distribution of levels of research within the programs. For this latter analysis, we assigned a project to one level of research according to the predominance of the level of research, using the KH Coder machine-learning tool. The resulting distribution throughout of the different financing channels is shown in Figure 5. The distribution is in accordance with the analysis of coded terms in Figure 4, but provides a more quantitative presentation of the main orientation within the programs and more direct insight in the time evolution.

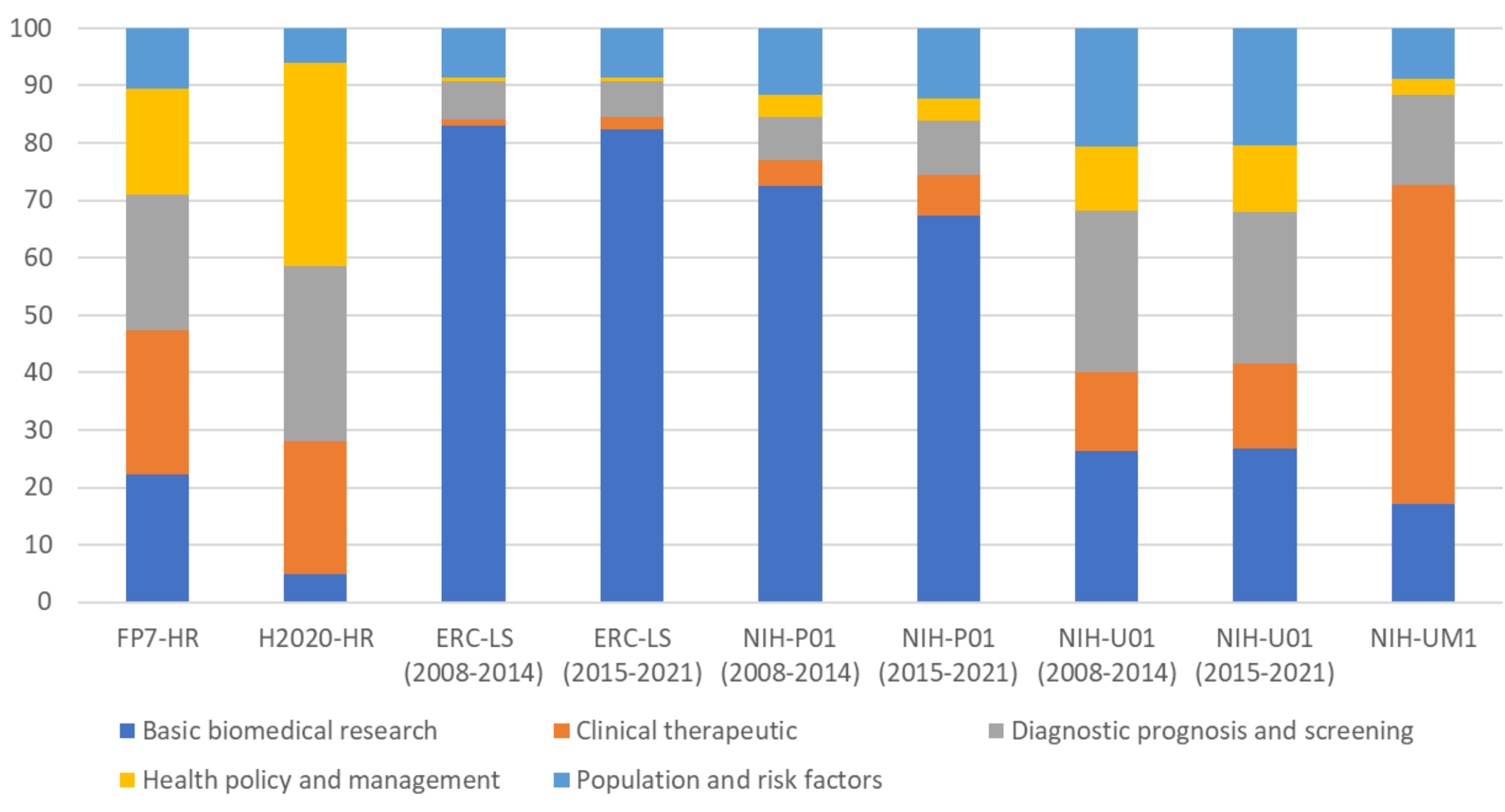

**Figure 5.** *Percentage distribution of projects classified through machine learning by health research levels in NIH-U01, NIH-P01, UM1 ERC-LS and FP7-HR and H2020-HR*

On the left hand side, the FP-HR funding shows the change in content over time. FP7-HR has projects distributed almost equally between the research levels, though with low fraction of research on population and risk factors. In contrast, in the H2020-HR program, more than a third of the projects are classified in the area of health management and policies, followed by diagnostics-prognosis-screening research with 30 percent, while less than 5 percent are basic biomedical projects. Next, data from ERC-LS shows that more than 80 percent of the ERC-LS projects are biomedical research projects, and that this fraction stays constant over time. Similarly to ERC-LS, the majority of projects funded by NIH-P01 are basic biomedical research (just over 72 percent in the 2008-2014 period and 67 percent in the 2015-2021 period) followed by population research projects and risk factors. Projects funded by NIH-U01 have a relatively balanced distribution among the different levels of research, although with a modest proportion of research projects on health management and policy (11 percent in both periods). Both NIH-PO1 and NIH-UO1 are stable over time. In contrast, in the more recent NIH-UM1 funding program, more than 55 percent of the projects were classified as clinical-therapeutic research, while a third of the projects were divided between basic biomedical research and diagnostics-prognosis-screening research. The remaining 10 percent of the projects were divided between population research and health management and policies.

## 3.3 Disease orientation of funding

To address the question whether project funding evolved in a disease and needs-driven perspective, we looked across the funding programs to assess in how far they included research related to three major disease areas: cancers, infectious diseases and cardiometabolic diseases. This analysis also aimed to inform whether specific instruments would be more directed at one or other disease area. The resulting heatmap, clustering funding streams according to the distribution of disease-coded terms, is presented in Figure 6.

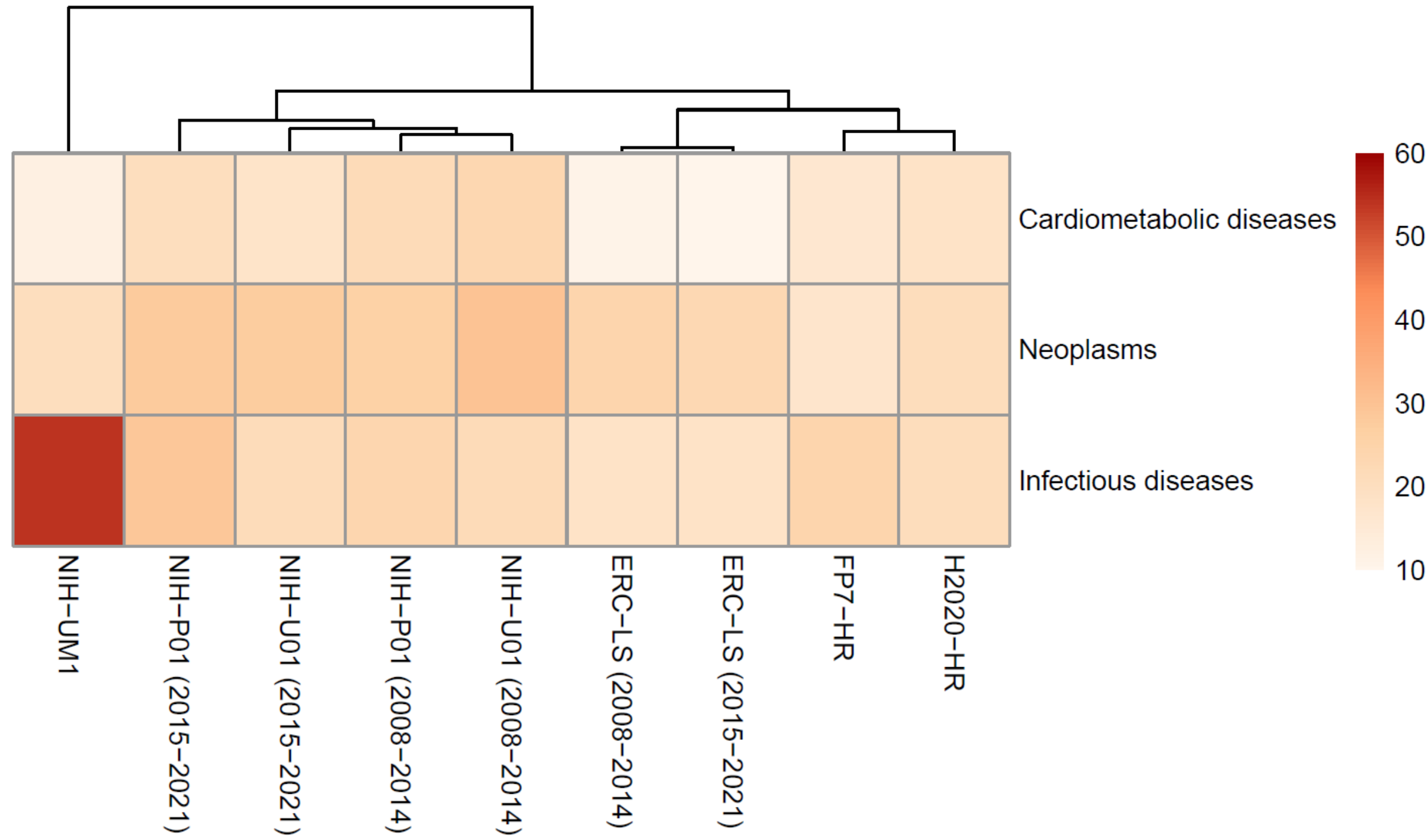

**Figure 6.** *Heat map resulting from clustering of funding programs according to the presence of terms related to three major diseases areas. The color gradient indicates the percentage of projects that contain in their description terms coded in one of the three defined disease groups.*

In this heatmap, UM1 is considered a single cluster because of its strong focus on infectious diseases. After setting aside UM1, interestingly, the higher level of clustering groups the US and European funding mechanisms separately suggesting different priorities for disease areas in health research between the NIH and the European Commission. Both FP7-HR and H2020-HR and the two periods of the ERC-LS are characterized by having a low percentage of projects with terms related to any of the three groups of diseases. In the case of ERC-LS, the terms related to cancer research are somewhat more prevalent, in FP7-HR terms related to infectious diseases and in H2020-HR, cancer and infectious disease in equal amounts. In NIH funding, excluding UM1, all diseases terms are more prevalent than in EU funding, with a stronger presence of cancer research.

When we examine the heatmap from the perspective of disease areas, it is clear that infectious disease is well represented and at the core of UM1 funding. Infectious disease is also present in the most recent group of P01 projects and to some extent in other programs. Cancer research on the other hand is present across all programs with only FP7 projects having less than 30% terms associated with cancer research. Cardiometabolic disease is less present, with most notably less than 20% of ERC projects having terms related to cardiometabolic disease.

In a final analysis, we examined the level of research within these disease areas, and performed a cross-tabulation between the five levels of research as classified by ML and the three disease groups. Figure 7 presents in each cell per disease area the frequency of terms related to the level of research, with a color-code for Pearson residuals that compares observed frequencies to expected ones.

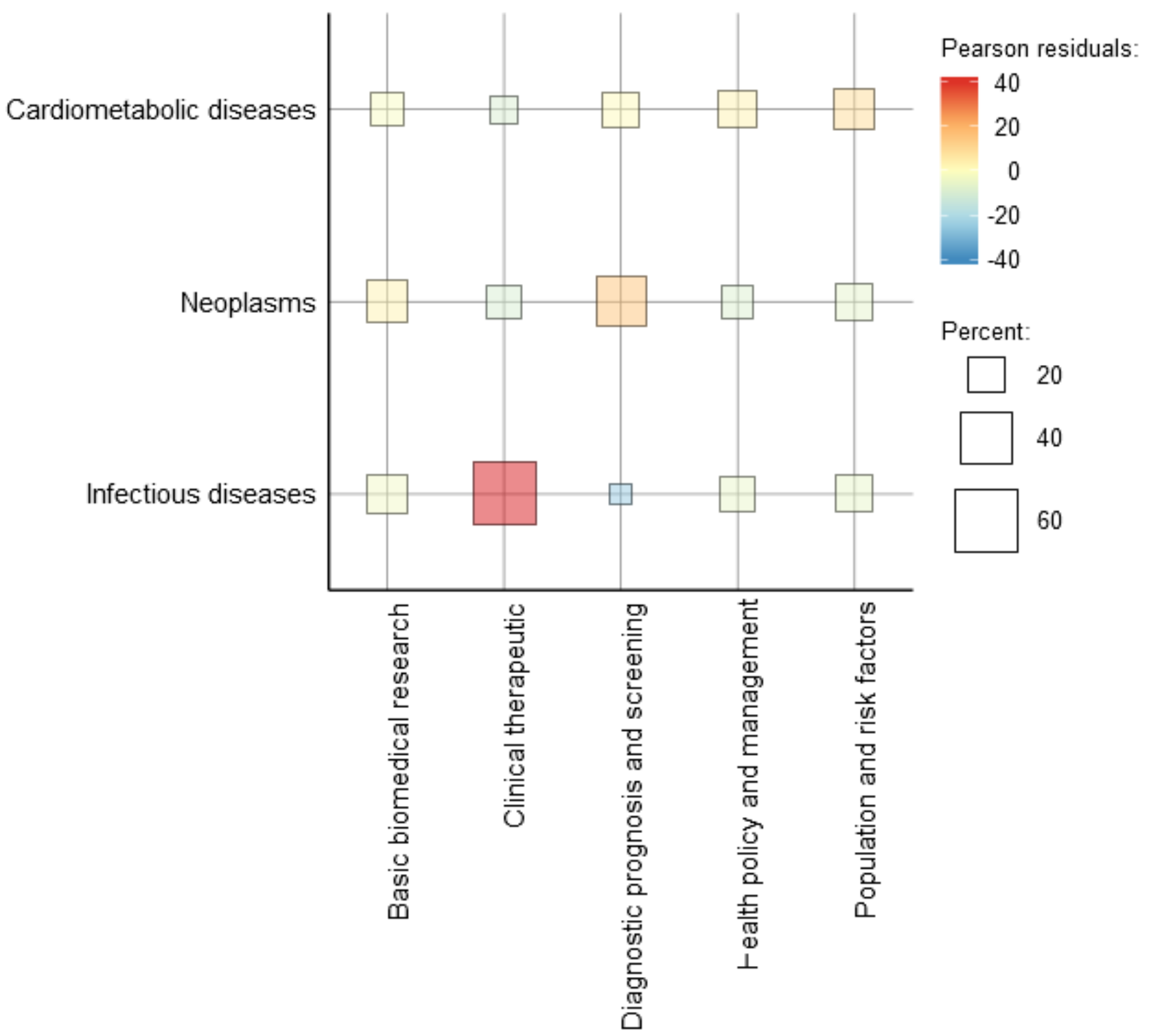

**Figure 7.** *Map of cross-tabulation of research levels vs disease groups indicating frequency of terms and Pearson residuals. In this map we compare the distribution of all the projects analyzed (all funding mechanisms) classified in the five levels of health research versus the distribution of the terms associated with the disease groups.*

The strongest presence is clinical therapeutic research within infectious disease (over 60 percent of projects), and diagnostics, prognosis and screening research within cancer research (40 percent of projects). Population and risk factors feature most prominently in research projects on cardiometabolic disease.

## 3.4 Evolution of the framework programs from FP7 to Horizon Europe

Our initial hypothesis of high flexibility in the projects with EU funding appears to be supported by the data above, i.e. in the disparate clustering of FP7-HR and H2020-HR in Figure 4, as well as in the content analysis using classification in Figure 5, with clearly different classification of projects within FP7-HR and H2020-HR. We used the data available for the first three years of Horizon Europe to confirm or refute this continuous adaption of the FP funding. As seen in Figure 8 of the distribution of terms coded by levels, there is a further transition towards population and risk levels, and particularly towards health management and policies, of among the successive FP health research projects. At the same time, the biomedical research projects maintain their initial decrease.

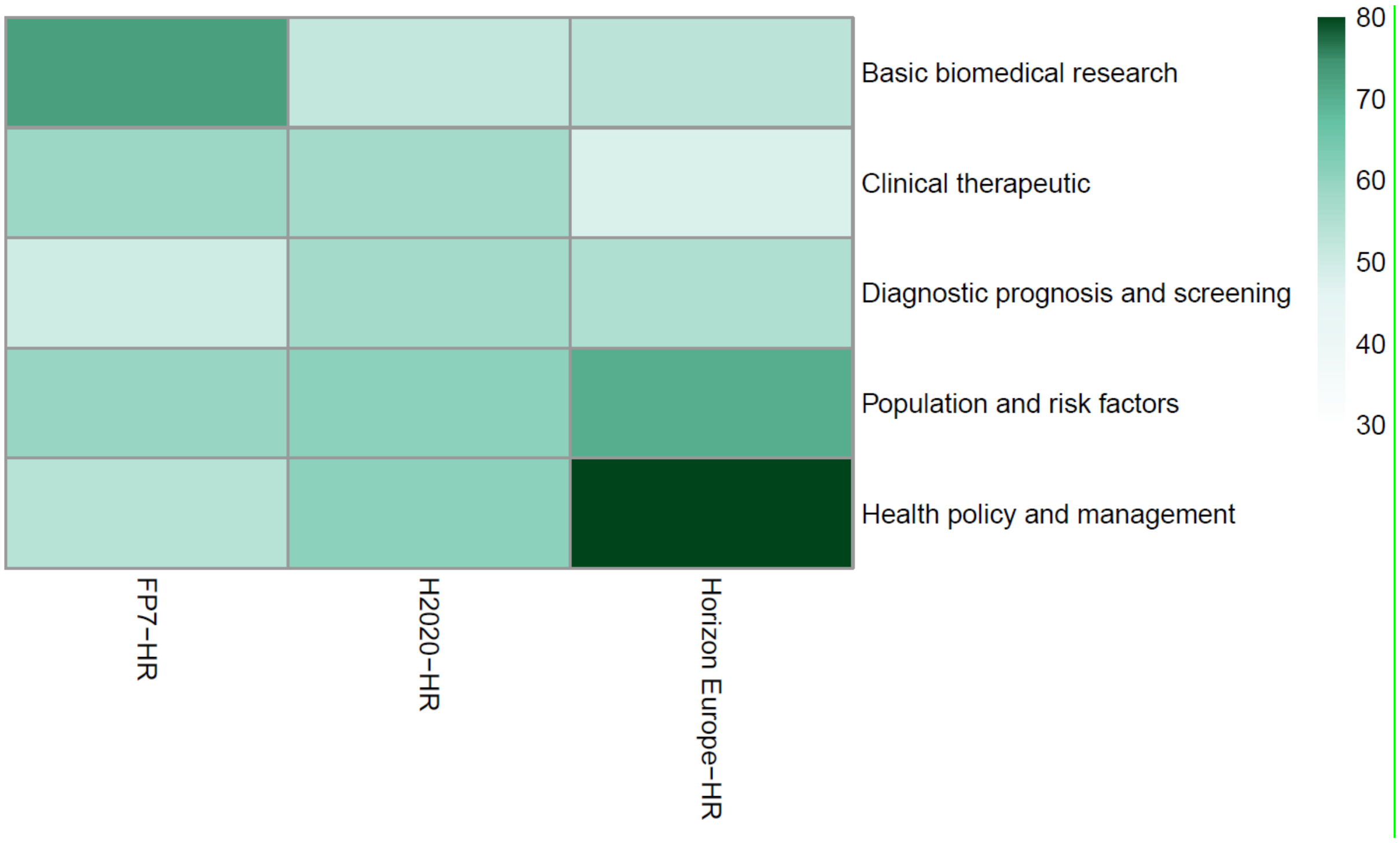

**Figure 8.** *Heat map of collaborative health research under FP7, H2020 and Horizon Europe according to the distribution of terms in health research levels*.

The distribution of projects with associated disease group terms further illustrates changes in the funding priorities with Horizon Europe (Figure 9). While the percentage of projects associated with infectious diseases remains prominent, the percentage of projects associated with neoplasms has grown steadily. Cardiometabolic disease shows a diminishing presence.

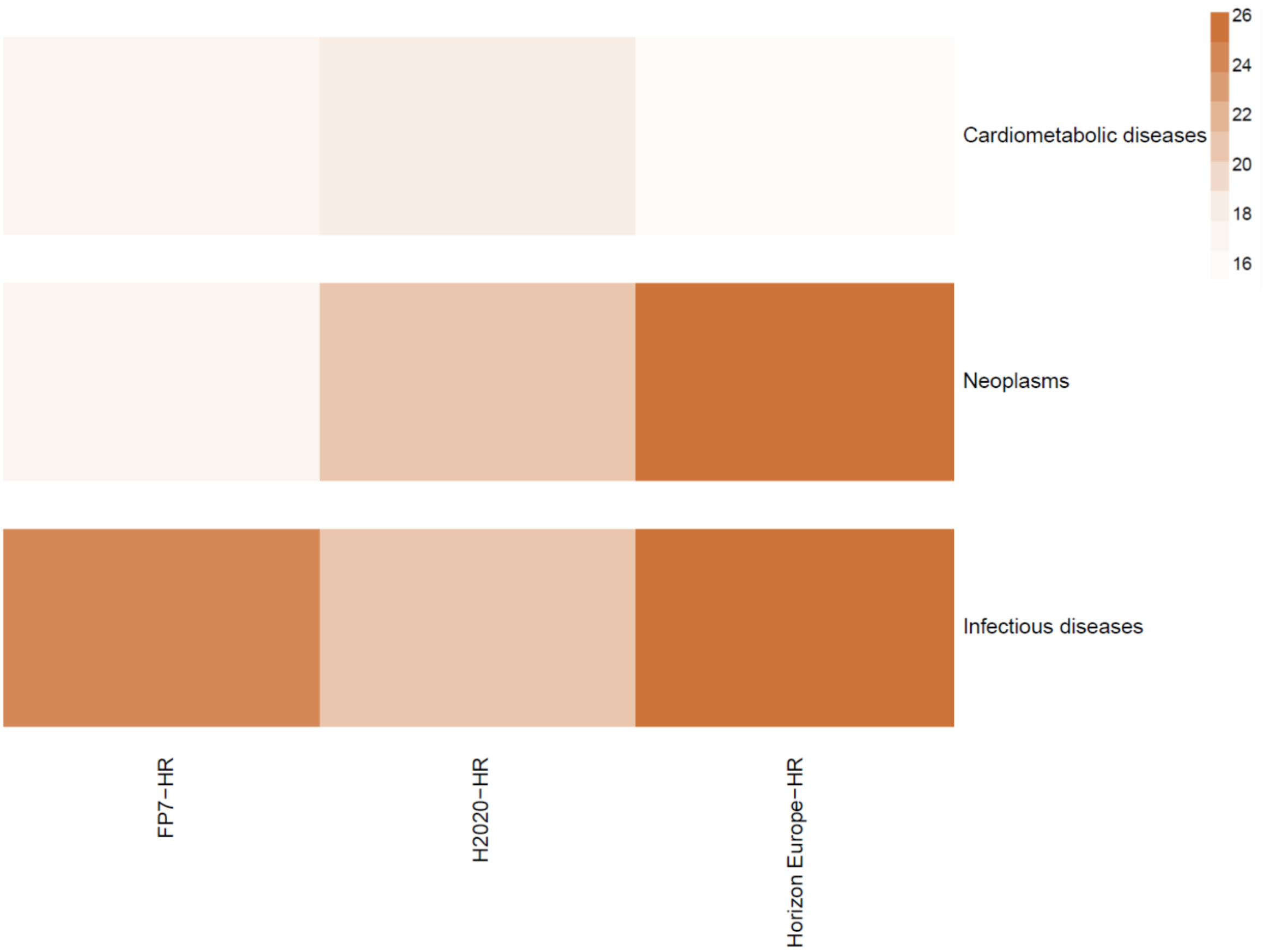

**Figure 9.** *Heat map of collaborative health research under FP7, H2020 and Horizon Europe according to the distribution of terms related to disease areas.*

Next, we sought to gain further details in the content of the research funded across the different FP-HR programs, using mapping of co-occurrence of terms (Figure 10). Although the central object of study of the last three framework programs revolves around the patient and the disease, the mapping further illustrates the differences in topics between them. Consistent with the analysis of the distribution of coded terms and with the classification of projects by research levels, FP7 shows an affinity to biomedical terms such as "cell," "mechanism," "drug" and "target." FP7 shares with H2020 terms related to a therapeutic clinical approach with a strong association to terms such as "therapy," "treatment" and "quality." H2020, for its part, is particularly associated with terms at involving clinical implementation level, and with terms related to technological management and innovation. H2020 and Horizon Europe share terms at the population level, however Horizon Europe is clearly specialized at the management and health policy level with terms such as "policy," implementation," "stakeholder," "guideline" and "decision."

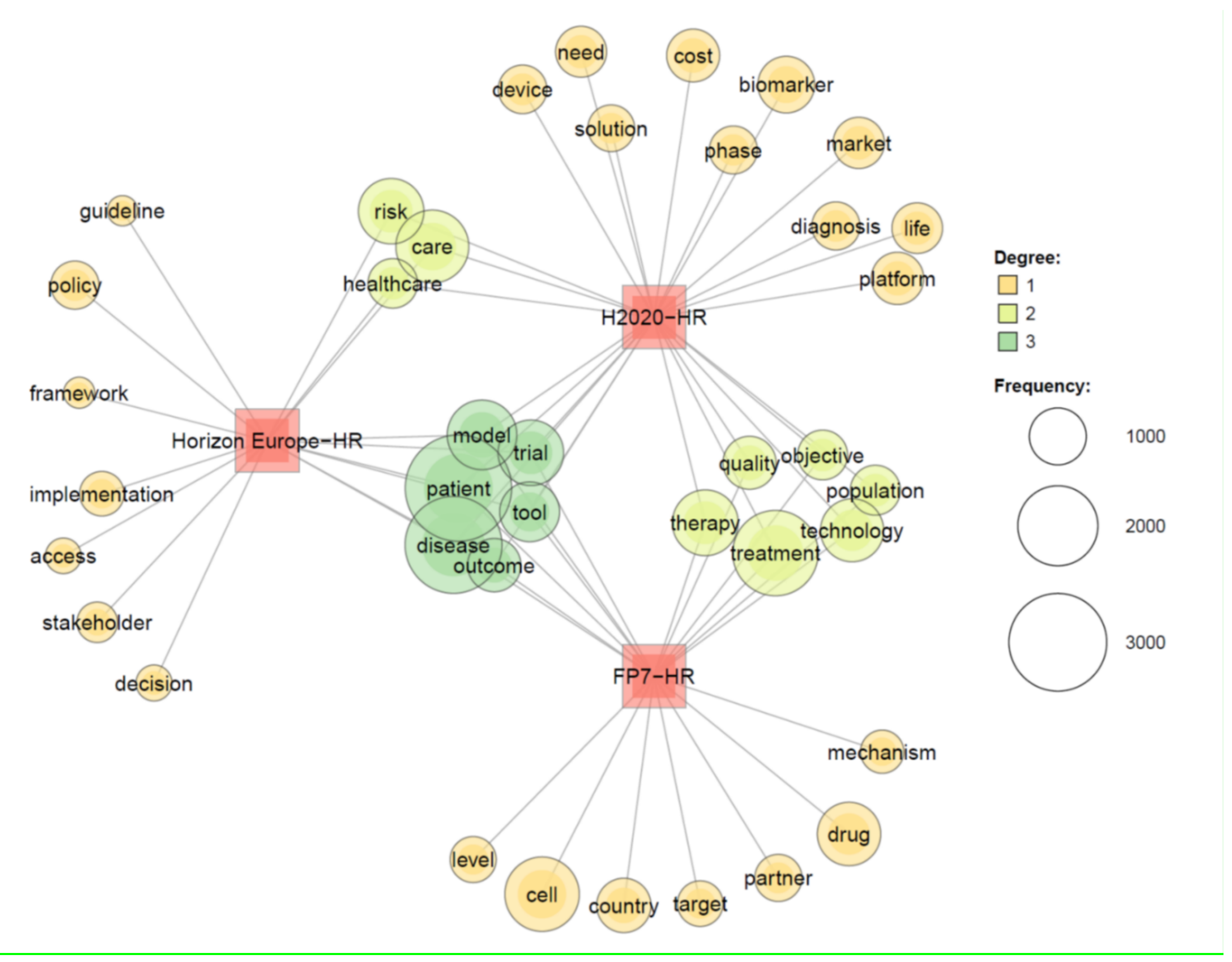

**Figure 10**. *Bimodal map of co-occurrence of terms in collaborative health research projects funded through FP7, H2020 and Horizon Europe.*

## 3.5 Publication-level analysis

### *3.5.1 Distribution of research-level terms in publications*

To examine the thematic orientation of research outputs, we first analyzed the distribution of research-level–specific keywords in the titles and abstracts of publications associated with each funding mechanism and programme period. Figure 11 presents a hierarchically clustered heat map of keyword distributions, allowing for direct comparison across funding instruments and over time.

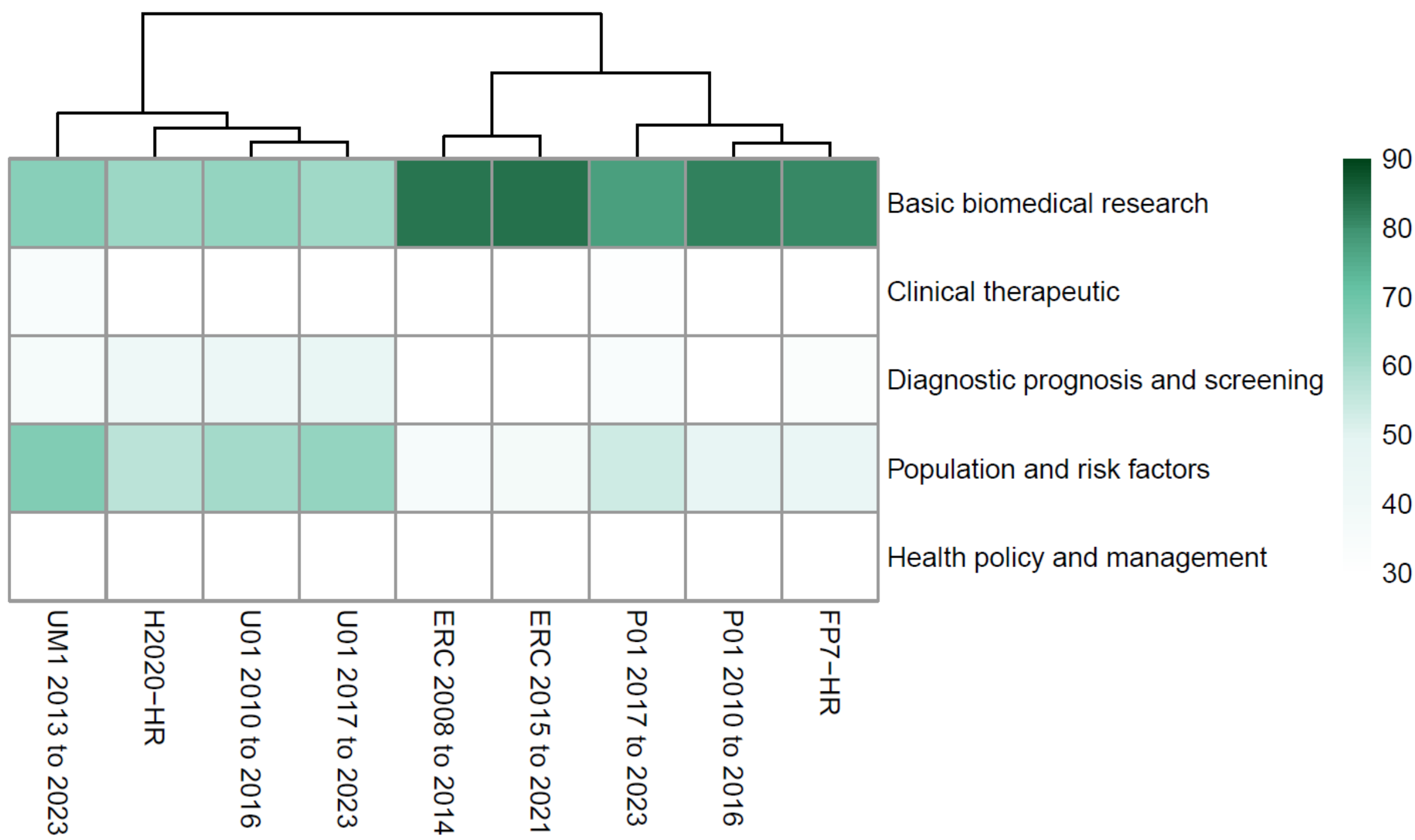

**Figure 11. Distribution of research-level keywords in publications.** *Heat map showing the relative frequency of research-level–specific keywords in the titles and abstracts of publications associated with EU FP-HR, ERC Life Sciences grants, and NIH funding mechanisms across programme periods. Hierarchical clustering highlights similarities and differences in thematic orientation across funding instruments.*

Across all funding mechanisms, publications exhibit a strong presence of terms associated with basic biomedical research, even in programmes where project-level analyses indicate a shift toward more applied or population-oriented research. This pattern is particularly pronounced for ERC-LS publications, where the dominance of basic biomedical terms remains stable across both programme periods. Similarly, NIH-funded publications associated with P01, U01, and UM1 mechanisms show consistently high frequencies of biomedical terms, with only modest variation over time.

In contrast, publications associated with the EU Framework Programmes display a more pronounced increase in terms related to population and risk factor research, diagnostic, prognostic, and screening activities, and health policy and management. Notably, while FP7 publications already exhibit a mixed profile, H2020 publications show a stronger relative presence of population-level and health systems–oriented terms, reflecting the broader societal challenge framing of health research in that programme.

Overall, the heat map reveals that while shifts in funding priorities are reflected in the lexical composition of publications, these changes are systematically more moderate than those observed at the project level. The persistence of biomedical terminology across all funding mechanisms suggests continuity in the foundations of health research outputs despite differences in funding orientation.

### *3.5.2 Classification of publications by research level*

Complementing the keyword-based analysis, publications were classified into five research levels using the supervised classification model described in the Methods section. Figure 12 summarizes the distribution of publications across research levels for each funding mechanism and programme period.

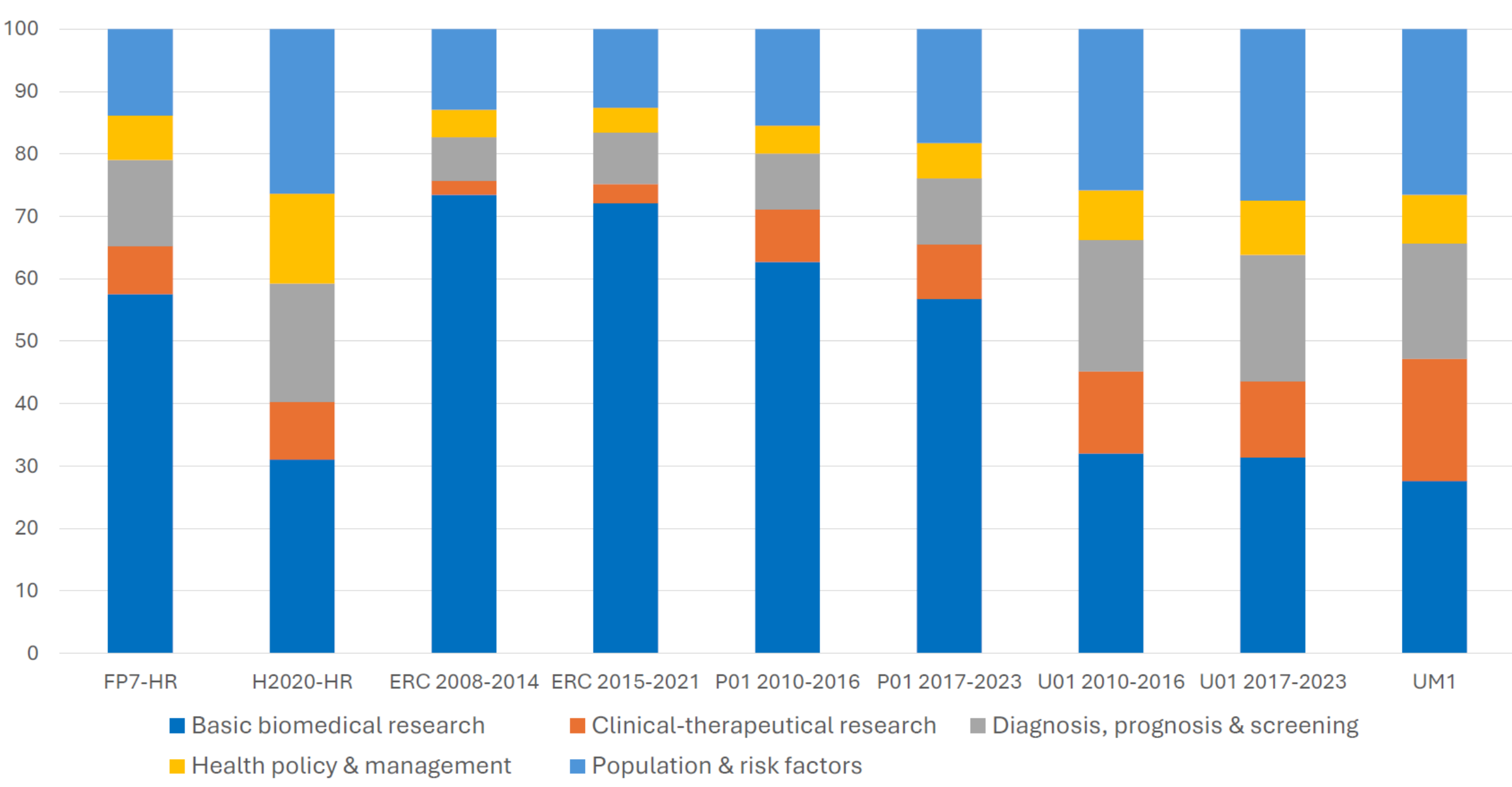

**Figure 12. Classification of publications by research level.** *Bar plot showing the distribution of publications across five research levels for EU FP-HR, ERC Life Sciences grants, and NIH funding mechanisms across programme periods, based on supervised classification of publication titles and abstracts.*

Consistent with the content analysis, basic biomedical research constitutes the largest share of publications across all funding mechanisms. This is particularly evident for ERC-LS publications, where more than 70% of outputs are classified as basic biomedical research in both programme periods, with only marginal variation over time. NIH-funded publications associated with P01 mechanisms also show a predominance of biomedical research, although a modest increase in population and risk factor research is observed in the later period.

Publications linked to U01 and UM1 mechanisms display a more balanced distribution across research levels, with higher shares of clinical, diagnostic, and population-oriented research compared to ERC and P01 outputs. Nevertheless, even in these mechanisms, basic biomedical research remains a central component of the publication portfolio.

For the EU Framework Programmes, the classification results reveal a marked shift between FP7 and H2020. While basic biomedical research dominates FP7 publications, its relative share decreases substantially in H2020, accompanied by increases in population and risk factor research, diagnostic and screening research, and health policy and management. Importantly, however, the magnitude of this shift is smaller than that observed at the project level, indicating partial rather than complete translation of funding priorities into research outputs.

Taken together, the classification results confirm that strategic funding policies influence the orientation of research outputs, but that these effects are attenuated relative to changes observed in funded project portfolios.

# 4 Discussion

## 4.1 Evolution towards higher levels of health research in the European FPs

The successive regulations of the EU FPs have increasingly emphasized the need to address health problems of the European population, most pronounced for Horizon Europe (European Commission Directorate-General for Research Innovation, 2021). This shift is a political decision, the result of the negotiation and planning process within the EU institutions of European Commission, Council representing Member States, and Parliament (Kim & Yoo, 2019; Ludwig et al., 2022; Sipido et al., 2022). Our data show that implementation of this policy in the call-based collaborative health program funding indeed has led to the funding of projects that are closer to translation into health care. Between FP7, Horizon 2020 and Horizon Europe, the EU funding for research under the strategic destination Health has visibly moved towards a predominance of higher levels of research. Of particular note is the expansion of research projects on health management and policies. The expectation is that these projects will result in short-term societal impact and impact-analysis has taken a more prominent position in the evaluation of the FPs (European Commission Directorate-General for Research and Innovation, 2023). In contrast to the collaborative call-based funding for health research, the ERC Life Sciences bottom-up projects have maintained a predominant focus of basic biomedical research.

While the project-level analysis provides clear evidence of a strategic reorientation of the European Framework Programmes, the publication-level results indicate that this shift is only partially reflected in research outputs. Publications associated with FP7 already exhibit a mixed profile, combining strong biomedical components with population-level and diagnostic research. In H2020, this profile evolves further, with an increased presence of publications focused on population and risk factor research, diagnostics, and health policy and management. However, the magnitude of this change is consistently smaller than that observed at the level of funded projects. Basic biomedical research remains a substantial component of the publication portfolio, even as it declines sharply in project descriptions.

This attenuation between project orientations and publication outputs suggests that the evolution of the Framework Programmes toward higher levels of health research is more pronounced in terms of funder expectations and project selection than in the realized content of scientific publications. Rather than indicating a lack of responsiveness to strategic priorities, this pattern reflects structural features of health research production, including the cumulative nature of scientific discovery, the interdependencies between basic and applied research, and publication incentives that continue to favor foundational biomedical knowledge. From this perspective, the transition from FP7 to H2020 can be understood as a reconfiguration of research agendas that reshapes the balance of funded activities without fully displacing the core of biomedical research.

## 4.2 Trend differences between the EU FPs and NIH funding

Our analysis of NIH funding finds that within collaborative programs such as NIH-U01 higher levels of research are funded along with biomedical research, and that this mixed level remains constant over time. NIH-P01 is focused mostly on biomedical research and this NIH focus remains constant over time as well. Nevertheless also within NIH funding there is a trend towards higher levels of research. UM1 is the largest NIH funding scheme, with a singular focus on clinical therapeutic research and infectious disease that started in 2011 and has seen considerable growth since. Notably as well, funding into the higher-level scheme of U01 has increased over the years, with less investment into P01, which funds predominantly basic biomedical research.

Although investigation of continuity of funding was not within the aims of the study, the NIH database includes information about whether awards are a continuation or a new application, and terms in the project description can likewise indicate continuity of topic, including in UO1. These data suggest opportunities for continuity of high-level research within the NIH funding scheme. For EU funding, in contrast, given that projects are awarded through targeted calls, researchers are less likely to have a continued funding of the same research topic.

Differences between NIH and EU also emerge in the programs funding biomedical research. The ERC granting scheme of the EU is unique in its scale of funding to a single PI for risk-taking research and vastly exceeds NIH funding through single PI schemes. In contrast, NIH supports biomedical research in collaborative projects. Another NIH strategy for fostering ground-breaking research lies in providing a longer time window of funding. NIH has two dedicated programs for long-term funding for outstanding investigators, R35(NIH National Institute of General Medical Sciences, 2024) and R37(NIH National Cancer Institute, 2020), mostly basic biomedical research. In contrast to ERC, the funding is mostly determined on merit, rather than project-based. This comparison, though preliminary, opens an interesting window on different strategies for investment in basic biomedical research.

The publication-level analysis reinforces this contrast while also qualifying its implications. Across NIH funding mechanisms, including P01, U01, and UM1, publications display a relatively stable distribution across research levels over time, with basic biomedical research remaining a dominant component. Although modest increases in population-level and diagnostic research are observed—particularly for U01 and UM1 mechanisms—the overall profile of NIH-funded publications changes less dramatically than that observed for EU Framework Programme publications. This stability mirrors the project-level results and reflects the continued centrality of investigator-initiated research within the NIH funding system, even in the presence of evolving strategic priorities.

The extension of the analysis to include publications derived from funded projects also responds to longstanding concerns in the literature regarding the relationship between funding priorities and research trajectories. Funding instruments may shape how research is framed and articulated at the proposal stage, while the subsequent evolution of scientific work may follow partially independent dynamics (Gläser & Laudel, 2016). Recent empirical work further highlights how research outcomes can exhibit degrees of divergence or unexpectedness relative to initial funding orientations (Aslan et al., 2024). By comparing project-level classifications with the thematic content of resulting publications, our analysis explicitly engages with this issue, allowing us to assess the extent to which portfolio-level funding orientations are reflected, transformed, or attenuated in research outputs.

In summary, both funding systems display an orientation toward higher-level health research, but they pursue this goal through distinct institutional approaches. In the EU Framework Programmes, shifts in research orientation are closely tied to the seven-year political and programmatic cycle, resulting in pronounced reconfigurations of funded project portfolios following changes in policy objectives. NIH funding, by contrast, is characterized by greater continuity within established funding mechanisms, with strategic priorities being advanced through the introduction of new programmatic opportunities alongside a strong emphasis on investigator-initiated research. While these approaches differ in their modes and degrees of steering, both systems converge in their articulation of research agendas aimed at addressing major health challenges. At the same time, the publication-level analysis indicates that such convergence is realized only partially in research outputs, underscoring that policy steering reshapes research portfolios within enduring epistemic and institutional constraints.

## 4.3 Evolution and EU-NIH differences in priority setting for disease areas

The terms analysis of content of the projects funded, did not lead to large, disease-related clusters in the maps, except for the identification of cancer research and infectious-disease related research. Interestingly, a disease-based clustering of funding separated EU and NIH funding with overall stronger presence of some specific disease terms in NIH funding. A likely explanation is the NIH strategy of funding through institutes (Supplementary information).

For many years the US has engaged for investment in diseases where need was considered the highest with notable examples of HIV research (Padamsee, 2018) and cancer (Geiger et al., 2024; Mervis, 2023). EU funding has similarly identified cancer as a major area of need, launching a Mission Cancer in Horizon Europe and a Beating Cancer Plan. With regard to infectious diseases, antimicrobial resistance and more recently Covid-19 have been a major driver for investment in research for all agencies across the world (Larrue, 2021; Sipido et al., 2020).

The data are in line with some concerns about underinvestment in cardiovascular disease. Such concerns have referred to the lack of new treatments in relation to the growing burden of cardiovascular disease (McClellan et al., 2019; Nicholls, 2018). A publication analysis also pointed in the same direction (Rafols et al., 2022; van de Klippe et al., 2022). Although NIH has increased the budget for the National Heart Lung and Blood Institutes (Supplementary Information), our data could not detect a substantial increase of projects under EU funding. Analysis of different levels in the present study points towards larger investment in preventative research than in discovery research. This would continue the trend noted in a publication analysis for the years up to 2013 which noted particular growth in the area of population sciences (Gal et al., 2019).

## 4.4 Virtues and limitations of the study

The scope of the study was restricted to core funding that primarily serves public institutions and could be analysed for trends over a longer period. The two health research funding agencies were chosen for the analysis because of their leading role in the global landscape, their strategic approach guided by governmental decisions, and accessible documentation. Future analysis would usefully look into other funding agencies with large financial impact, including charities such as Wellcome with a budget close to that of EU health research investment. Within the NIH budget, we excluded the largest funding scheme being the R01 grants. This choice was motivated by the nature of R01 funding, i.e. scientist-driven and mostly led by single PI. The NIH budget we considered, nevertheless is considerably larger than the budget invested by EU. Indeed, the largest spending in Europe is through national funding agencies. However, with some exceptions, the data at national level for the analysis performed here are difficult to obtain. It is noteworthy that European national funding mechanisms have also initiated funding schemes to promote translation and innovation, and collaborative research. Moreover, because the analysis operates at the level of multi-year FP programme periods, incremental adjustments within programmes may not be fully detected, particularly when they concern narrower subtopics rather than overarching thematic and research-level orientations.

The methodology and approach has a number of inherent limitations that must be acknowledged. First, both project descriptions and publication texts reflect intentions, framings, and reporting practices rather than the full complexity of research activities as they unfold over time. Grant proposals are strategic documents, shaped by researchers' efforts to align with funding priorities, while publications represent selective outputs influenced by publication incentives and disciplinary norms. The divergence observed between project-level and publication-level results should therefore be interpreted not as measurement error, but as an inherent feature of research systems in which intentions and outcomes are only partially aligned. Although the same analytical framework was applied to both datasets, project descriptions and scientific publications are distinct document types with different communicative purposes, temporal orientations, and stylistic conventions. These genre differences may themselves influence thematic signals and classification patterns, and thus should be considered when interpreting comparisons between project-level and publication-level results.

Second, the classification of research into broad levels and disease categories involves trade-offs between analytical clarity and granularity. While the use of coarse categories enables cross-programme and longitudinal comparison, it may obscure finer-grained shifts within subfields or disease areas, as well as boundary dynamics between adjacent categories. The purpose of the present analysis is not to exhaustively map all within-category variation, but to identify macroscopic shifts in the orientation of research portfolios that are visible across programmes and funding models. More fine-grained distinctions—such as differentiating sub-areas within major disease groups or distinguishing between priority and neglected conditions—represent promising directions for future research building on the framework developed here. Similarly, although the combination of keyword-based content analysis and supervised classification increases robustness, no classification scheme can fully capture the multidimensional and evolving nature of health research. In particular, because projects often span multiple research levels, classification necessarily reflects the dominant orientation of a project rather than its full cross-level complexity, although sensitivity analyses indicated that alternative thresholding did not materially affect aggregate programme-level patterns. Future work could build on this approach by incorporating more granular classifications, alternative machine-learning methods, or complementary qualitative analyses to further refine understanding of how funding priorities shape research trajectories.

A central virtue of the present study lies in its portfolio-level perspective, which allows for systematic comparison across funding mechanisms and programme periods using consistent analytical tools. By focusing on funding mechanisms rather than individual projects or researchers, the analysis captures macroscopic shifts in research orientation that are often difficult to detect in case-based or micro-level studies. The parallel analysis of project descriptions and resulting scientific publications further strengthens this perspective by enabling an explicit comparison between funder expectations, as expressed through funded projects, and the realized orientation of research outputs. Our study complements existing work on funding priorities and disease allocations by shifting attention from specific thematic distributions to the relationship between funding mechanisms, project orientations, and research outputs.

Several previous studies on financing for health research tended to focus on the analysis of beneficiary institutions and researchers (Hoppe et al., 2019; Madsen & Aagaard, 2020; Ross et al., 2022) and on the final products of these projects, such as publications and patents (Fajardo-Ortiz et al., 2022; Mugabushaka et al., 2022). As such, our study addresses a knowledge gap regarding which research topics (groups of diseases) are financed and from what focus or level of research (basic, clinical, population-based, and management and health policies). The strength compared to publications research, is the short time lag after funding decision, with publications following only with several years of delay. In addition, analysis of project content provides a direct link compared to uncertainty when linking publications to funding. The information gathered tells about the funders' aims and expectations, with the publication analysis giving insight in the first output level though it is quite possible that the outcomes eventually will diverge from the funding aims, a topic for future research.

# 5 Conclusions

Our findings contribute to ongoing debates on the elasticity of science and the capacity of strategic funding policies to redirect research trajectories. They suggest that while large-scale funding programmes can reshape research portfolios at the level of funded projects, their influence on the production of scientific knowledge may be more constrained and indirect. By comparing both project descriptions and the content of resulting scientific publications, our analysis highlights that while strategic funding policies clearly reshape funder expectations and funded portfolios, their effects on realized research outputs are more moderate, reflecting structural constraints in the production of health knowledge.

# 6 Supporting information

Supplementary information: EU and NIH funding policies and strategic planning; Methodology for coding of terms by level of research; Co-occurrence maps of NIH funding mechanisms.

# 7 Competing Interests

The authors declare that they have no competing interests.

# 8 Data Availability Statement

The data underlying this article are available in the article and in its online supplementary material.

# 9 Funding

The authors did not receive specific funding for this research.

# 10 Author Contributions (CRediT taxonomy)

David Fajardo-Ortiz: Conceptualization; Methodology; Formal analysis; Writing – original draft; Writing – review & editing.

Karin R. Sipido: Conceptualization; Supervision; Writing – original draft; Writing – review & editing.

Bart Thijs: Methodology; Writing – review & editing.

Wolfgang Glänzel: Methodology; Writing – review & editing.

# 11 References

Aarden, E., Marelli, L., & Blasimme, A. (2021). The translational lag narrative in policy discourse in the United States and the European Union: a comparative study. *Humanities and Social Sciences Communications*, *8*(1), 1-9.

Al, M., Levison, S., Berdel, W. E., & Andersen, D. Z. (2023). Decentralised elements in clinical trials: recommendations from the European Medicines Regulatory Network. *Lancet*, *401*(10385), 1339.

Arnold, E., & Barker, K. E. (2022). What past changes in Swedish policy tell us about developing third-generation research and innovation governance. In *Smart Policies for Societies in Transition* (pp. 59-86). Edward Elgar Publishing.

Aslan, Y., Yaqub, O., Sampat, B. N., & Rotolo, D. (2024). Unexpectedness in medical research. *Research Policy*, *53*(8), 105075. https://doi.org/https://doi.org/10.1016/j.respol.2024.105075

Ballreich, J. M., Gross, C. P., Powe, N. R., & Anderson, G. F. (2021). Allocation of National Institutes of Health Funding by Disease Category in 2008 and 2019. *JAMA Network Open*, *4*(1), e2034890-e2034890. https://doi.org/10.1001/jamanetworkopen.2020.34890

Bowen, A., & Casadevall, A. (2015). Increasing disparities between resource inputs and outcomes, as measured by certain health deliverables, in biomedical research. *Proceedings of the National Academy of Sciences*, *112*(36), 11335-11340. https://doi.org/10.1073/pnas.1504955112

Bush, V. (1945). *Science, the endless frontier. A report to the President*. United States Government Printing Office.

Bush, V. (1947). A National Science Foundation: Statement before the Committee on Interstate and Foreign Commerce, House of Representatives, March 7, 1947. *Science*, *105*(2725), 302-305.

Caulfield, T. (2010). Stem cell research and economic promises. *Journal of Law, Medicine & Ethics*, *38*(2), 303-313.

Coburn, J., Yaqub, O., Ràfols, I., & Chataway, J. (2024). Cross-disease spillover from research funding: Evidence from four diseases. *Social Science & Medicine*, *349*, 116883. https://doi.org/https://doi.org/10.1016/j.socscimed.2024.116883

Demotes-Mainard, J., & Ohmann, C. (2005). European Clinical Research Infrastructures Network: promoting harmonisation and quality in European clinical research. *Lancet*, *365*(9454), 107-108.

European Commission Directorate-General for Research and Innovation. (2023). *Horizon Europe programme analysis - Impact assessment, evaluation and monitoring of Horizon Europe*. European Commission. https://research-and-innovation.ec.europa.eu/strategy/support-policy-making/shaping-eu-research-and-innovation-policy/evaluation-impact-assessment-and-monitoring/horizon-europe-programme-analysis_en#monitoring-horizon-europe

European Commission Directorate-General for Research Innovation. (2021). *Horizon Europe : strategic plan 2021-2024*. Publications Office of the European Union. https://doi.org/doi/10.2777/083753

Fajardo-Ortiz, D., Hornbostel, S., de Wit, M. M., & Shattuck, A. (2022). Funding CRISPR: Understanding the role of government and philanthropic institutions in supporting academic research within the CRISPR innovation system. *Quantitative Science Studies*, *3*(2), 443-456. https://doi.org/10.1162/qss_a_00187

Fajardo, D., & Castano, V. M. (2016). Hierarchy of knowledge translation: from health problems to ad-hoc drug design. *Current Medicinal Chemistry*, *23*(26), 3000-3012.

Gal, D., Thijs, B., Glänzel, W., & Sipido, K. R. (2019). Hot topics and trends in cardiovascular research. *European Heart Journal*, *40*(28), 2363-2374.

Geiger, A. M., Jaffee, E. M., Berger, M. S., Brown, C. L., Rathmell, W. K., & Bertagnolli, M. M. (2024). An orientation to the US National Cancer plan for the research community. *JNCI: Journal of the National Cancer Institute*, *116*(6), 789-794. https://doi.org/10.1093/jnci/djae037

Gillum, L. A., Gouveia, C., Dorsey, E. R., Pletcher, M., Mathers, C. D., McCulloch, C. E., & Johnston, S. C. (2011). NIH Disease Funding Levels and Burden of Disease. *PLOS ONE*, *6*(2), e16837. https://doi.org/10.1371/journal.pone.0016837

Gläser, J., & Laudel, G. (2016). Governing Science: How Science Policy Shapes Research Content. *European Journal of Sociology*, *57*(1), 117-168. https://doi.org/10.1017/S0003975616000047

Gross Cary, P., Anderson Gerard, F., & Powe Neil, R. (1999). The Relation between Funding by the National Institutes of Health and the Burden of Disease. *New England Journal of Medicine*, *340*(24), 1881-1887. https://doi.org/10.1056/NEJM199906173402406

Higuchi, K. (2016). KH Coder 3 reference manual. *Kioto (Japan): Ritsumeikan University*.

Hoffman, S. J., Røttingen, J.-A., Bennett, S., Lavis, J. N., Edge, J. S., & Frenk, J. (2012). Background paper on conceptual issues related to health systems research to inform a WHO global strategy on health systems research. *Health Systems Alliance*.

Hoppe, T. A., Litovitz, A., Willis, K. A., Meseroll, R. A., Perkins, M. J., Hutchins, B. I., Davis, A. F., Lauer, M. S., Valantine, H. A., & Anderson, J. M. (2019). Topic choice contributes to the lower rate of NIH awards to African-American/black scientists. *Science Advances*, *5*(10), eaaw7238.

Iwata, R., Kuramoto, K., & Kumagai, S. (2021). Extracting Potential Innovators Using Innovator Scores and Naive Bayes Classifier. 2021 IEEE 10th Global Conference on Consumer Electronics (GCCE),

Kastrinos, N., & Weber, K. M. (2020). Sustainable development goals in the research and innovation policy of the European Union. *Technological Forecasting and Social Change*, *157*, 120056.

Kim, J., & Yoo, J. (2019). Science and technology policy research in the EU: from Framework Programme to HORIZON 2020. *Social Sciences*, *8*(5), 153.

König, T. (2017). *The European research council*. John Wiley & Sons.

Larrue, P. (2021). Mission-oriented innovation policy in Norway. https://doi.org/doi:https://doi.org/10.1787/2e7c30ff-en

Lawler, M., De Lorenzo, F., Lagergren, P., Mennini, F. S., Narbutas, S., Scocca, G., Meunier, F., & European Academy of Cancer, S. (2021). Challenges and solutions to embed cancer survivorship research and innovation within the EU Cancer Mission. *Molecular Oncology*, *15*(7), 1750-1758. https://doi.org/10.1002/1878-0261.13022

Llewellyn, N., Carter, D. R., DiazGranados, D., Pelfrey, C., Rollins, L., & Nehl, E. J. (2020). Scope, influence, and interdisciplinary collaboration: the publication portfolio of the NIH Clinical and Translational Science Awards (CTSA) Program from 2006 through 2017. *Evaluation and the Health Professions*, *43*(3), 169-179.

Ludwig, D., Blok, V., Garnier, M., Macnaghten, P., & Pols, A. (2022). What's wrong with global challenges? *Journal of Responsible Innovation*, *9*(1), 6-27.

Luukkonen, T. (2014). The European Research Council and the European research funding landscape. *Science and Public Policy*, *41*(1), 29-43. https://doi.org/10.1093/scipol/sct031

Madsen, E. B., & Aagaard, K. (2020). Concentration of Danish research funding on individual researchers and research topics: Patterns and potential drivers. *Quantitative Science Studies*, *1*(3), 1159-1181. https://doi.org/10.1162/qss_a_00077

Madsen, E. B., & Andersen, J. P. (2024). Funding priorities and health outcomes in Danish medical research. *Social Science & Medicine*, *360*, 117347. https://doi.org/https://doi.org/10.1016/j.socscimed.2024.117347

Mazzucato, M. (2018). *Mission-oriented research & innovation in the European Union : a problem-solving approach to fuel innovation-led growth*. Publications Office. https://doi.org/doi/10.2777/360325

McClellan, M., Brown, N., Califf, R. M., & Warner, J. J. (2019). Call to Action: Urgent Challenges in Cardiovascular Disease: A Presidential Advisory From the American Heart Association. *Circulation*, *139*(9), e44-e54. https://doi.org/doi:10.1161/CIR.0000000000000652

Mervis, J. (2023). Biden backs science in 2024 spending blueprint. *Science*, *379*(6637), 1078. https://doi.org/10.1126/science.adh8293

Miller, K., McAdam, R., & McAdam, M. (2018). A systematic literature review of university technology transfer from a quadruple helix perspective: toward a research agenda. *R&D Management*, *48*(1), 7-24. https://doi.org/https://doi.org/10.1111/radm.12228

Moore, D. A. Q., Yaqub, O., & Sampat, B. N. (2024). Manual versus machine: How accurately does the Medical Text Indexer (MTI) classify different document types into disease areas? *PLOS ONE*, *19*(3), e0297526. https://doi.org/10.1371/journal.pone.0297526

Mugabushaka, A.-M., van Eck, N. J., & Waltman, L. (2022). Funding COVID-19 research: Insights from an exploratory analysis using open data infrastructures. *Quantitative Science Studies*, *3*(3), 560-582. https://doi.org/10.1162/qss_a_00212

Myers, K. (2020). The Elasticity of Science. *American Economic Journal: Applied Economics*, *12*(4), 103–134. https://doi.org/10.1257/app.20180518

Nathan, D. G., Fontanarosa, P. B., & Wilson, J. D. (2001). Opportunities for medical research in the 21st century. *JAMA*, *285*(5), 533-534.

Nederbragt, H. (2000). The biomedical disciplines and the structure of biomedical and clinical knowledge. *Theoretical Medicine and Bioethics*, *21*, 553-566.

Neta, G., Sanchez, M. A., Chambers, D. A., Phillips, S. M., Leyva, B., Cynkin, L., Farrell, M. M., Heurtin-Roberts, S., & Vinson, C. (2015). Implementation science in cancer prevention and control: a decade of grant funding by the National Cancer Institute and future directions. *Implementation Science*, *10*, 1-10.

Nicholls, M. (2018). Funding of cardiovascular research in the USA: Robert Califf and Peter Libby – speak about cardiovascular research funding in the United States and what the latest trends are with Mark Nicholls. *European Heart Journal*, *39*(40), 3629-3631. https://doi.org/10.1093/eurheartj/ehy638

NIH National Cancer Institute. (2020). *MERIT Award (R37)*. Retrieved 2024 from https://www.cancer.gov/grants-training/grants-funding/funding-opportunities/merit

NIH National Institute of General Medical Sciences. (2024). Maximizing Investigators' Research Award (MIRA) (R35). https://www.nigms.nih.gov/Research/mechanisms/MIRA

OECD. (2021). *OECD Science, Technology and Innovation Outlook 2021*. https://doi.org/doi:https://doi.org/10.1787/75f79015-en

Padamsee, T. J. (2018). Fighting an Epidemic in Political Context: Thirty-Five Years of HIV/AIDS Policy Making in the United States. *Social History of Medicine*, *33*(3), 1001-1028. https://doi.org/10.1093/shm/hky108

Pratt, B., Wild, V., Barasa, E., Kamuya, D., Gilson, L., Hendl, T., & Molyneux, S. (2020). Justice: a key consideration in health policy and systems research ethics. *Bmj Global Health*, *5*(4), e001942.

Rafols, I., Yegros, A., van de Klippe, W., & Willemse, T. (2022). Mapping Research on Cardio-Metabolic Diseases. *SocArXiv*. https://doi.org/10.31235/osf.io/zn5gf

Reed, J. C., White, E. L., Aubé, J., Lindsley, C., Li, M., Sklar, L., & Schreiber, S. (2012). The NIH's role in accelerating translational sciences. *Nature Biotechnology*, *30*(1), 16-19.

Roberts, M. C., Clyne, M., Kennedy, A. E., Chambers, D. A., & Khoury, M. J. (2019). The current state of funded NIH grants in implementation science in genomic medicine: a portfolio analysis. *Genetics in Medicine*, *21*(5), 1218-1223.

Rodríguez, H., Fisher, E., & Schuurbiers, D. (2013). Integrating science and society in European Framework Programmes: Trends in project-level solicitations. *Research Policy*, *42*(5), 1126-1137.

Ross, M. B., Glennon, B. M., Murciano-Goroff, R., Berkes, E. G., Weinberg, B. A., & Lane, J. I. (2022). Women are credited less in science than men. *Nature*, *608*(7921), 135-145. https://doi.org/10.1038/s41586-022-04966-w

Shaw, J. (2022). There and back again: Revisiting Vannevar Bush, the linear model, and the freedom of science. *Research Policy*, *51*(10), 104610.

Singer, D. S. (2022). A new phase of the Cancer Moonshot to end cancer as we know it. *Nature Medicine*, *28*(7), 1345-1347. https://doi.org/10.1038/s41591-022-01881-5

Sipido, K., Antoñanzas, F., Celis, J., Degos, L., Frackowiak, R., Fuster, V., Ganten, D., Gay, S., Hofstraat, H., Holgate, S. T., Krestin, G., Manns, M., Meunier, F., Oertel, W., Palkonen, S., Pavalkis, D., Rübsamen-Schaeff, H., Smith, U., Stallknecht, B. M., & Zima, T. (2020). Overcoming fragmentation of health research in Europe: lessons from COVID-19. *The Lancet*, *395*(10242), 1970-1971. https://doi.org/10.1016/S0140-6736(20)31411-2

Sipido, K., Fajardo-Ortiz, D., Vercruysse, T., Glänzel, W., & Veugelers, R. (2022). *Fostering coherence in EU health research: Strengthening EU research for better health*. European Parliament. https://doi.org/10.2861/711150

van de Klippe, W., Yegros, A., Willemse, T., & Rafols, I. (2022). Priorities in research portfolios: why more upstream research is needed in cardiometabolic and mental health. *SocArXiv*. https://doi.org/10.31235/osf.io/xrhgd

Viergever, R. F., & Hendriks, T. C. (2016). The 10 largest public and philanthropic funders of health research in the world: what they fund and how they distribute their funds. *Health Research Policy and Systems*, *14*(1), 1-15.

Yaqub, O., Coburn, J., & Moore, D. A. Q. (2024). Research-targeting, spillovers, and the direction of science: Evidence from HIV research-funding. *Research Policy*, *53*(8), 105076. https://doi.org/https://doi.org/10.1016/j.respol.2024.105076